\documentclass[amsmath,amssymb,superscriptaddress,showpac]{revtex4-2}
\usepackage{graphicx}
\usepackage{dcolumn}
\usepackage{amsmath}
\usepackage[latin1]{inputenc}
\usepackage{graphicx, psfrag}
\usepackage{amssymb}
\usepackage[colorlinks=true, citecolor=blue, urlcolor = blue, linkcolor= red, bookmarks=true]{hyperref}
\usepackage{float}
\usepackage{amsmath}
\usepackage{amsfonts}
\usepackage{dcolumn}
\usepackage{hyperref}
\usepackage{subfigure}
\usepackage{pgfplots}
\usepackage{epstopdf}

\makeatletter
\def\btt#1{\texttt{\@backslashchar#1}}
\DeclareRobustCommand\bblash{\btt{\@backslashchar}} \makeatother

\makeatletter
\def\btt#1{\texttt{\@backslashchar#1}}
\DeclareRobustCommand\bblash{\btt{\@backslashchar}} \makeatother

\begin{document}	
\newcommand{\CTP}{
\affiliation{Center for
		Theoretical Physics, Jamia Millia Islamia, New Delhi 110025, India\\Astrophysics and Cosmology Research Unit,
	School of Mathematics, Statistics and Computer Science,
	University of KwaZulu-Natal, Private Bag X54001,
	Durban 4000, South Africa}
}
\newcommand{\IIT}{\affiliation{Center for
		Theoretical Physics, Jamia Millia Islamia, New Delhi 110025, India\\
		Department of Physical Sciences, Indian Institute of Science Education and Research Kolkata, Mohanpur, Nadia, West Bengal, India
}}

\title[]{Thermodynamics and phase transition of rotating regular-de Sitter black holes}

\author{Md Sabir Ali}
\email{alimd.sabir3@gmail.com}
\IIT
\author{Sushant G. Ghosh}
\email{sghosh2@jmi.ac.in}
\CTP
\begin{abstract}
\noindent
We analyze thermodynamic properties of the rotating regular black holes having mass ($M$), angular momentum ($a$), and a magnetic charge $(g)$, and encompass Kerr black hole ($g=0$). The mass $M$ has a minimum at the radius $r_+=r_+^{\star}$, where both the heat capacity and temperature vanish. The thermal phase transition is because of the divergence of heat capacity at a critical radius $r_{+}^C$ with stable (unstable) branches for $r_+<r_+^C$ ($>r_+^C$).  We also generalize the rotating regular black holes in de Sitter (dS) background and analyzed its horizon structure to show that for each $g$, there are two critical values of the mass parameter $M_{\text{cr1}}$ and $M_{\text{cr2}}$ which correspond to the degenerate horizons. Thus, we have rotating regular-dS black holes with an additional cosmological horizon apart from the inner (Cauchy) and the outer (event) horizons. Next, we discuss the effective thermodynamic quantities of the rotating regular-dS black holes in the extended phase space where the cosmological constant ($\Lambda$) is considered as thermodynamic pressure. Combining the first laws at the two horizons, we calculate the heat capacity at constant pressure $C_P$, the volume expansion coefficient $\alpha$, and the isothermal compressibility $\kappa_T$. At a critical point, the specific heat at constant pressure, the volume expansion coefficient, and the isothermal compressibility of the regular-dS black holes exhibit an infinite peak suggesting a second-order phase transition.
\end{abstract}
\maketitle

\section{Introduction}
The gravitational collapse of a star with sufficient mass, under some general conditions, will result in a singularity--as stated by the singularity theorems because of Hawking and Ellis \cite{Hawking:1970, Hawking:1973}. The regular black holes, dating back to Bardeen \cite{Bardeen:1968}, avoid the curvature singularity beyond the event horizon \cite{Borde:1994ai, ABG}. Bardeen proposed the first regular model based on the ideas of Sakharov \cite{Sakharov:1966}, and Gliner \cite{Gliner:1966} who suggested that the appearance of the singularities may be avoided if the matter has a de Sitter (dS) core, i.e., with the equation of state $p=-\rho$. The Bardeen metric is spherically symmetric static, which is asymptotically flat, and has regular centers, satisfying the weak energy condition, and it has influenced the direction of subsequent research on the existence or avoidance of singularities. Later, the Bardeen model was interpreted as an exact solution to Einstein's field equations coupled to nonlinear electrodynamics (NED) \cite{ABG}. It is also contrary to the possibility that singularity avoidance may manifest in black hole spacetimes even if we ignore the strong energy condition \cite{Borde:1994ai, Borde:1996df}. The weak cosmic censorship conjecture \cite{Hawking:1970, Hawking:1973} asserts that an event horizon always surrounds the singularities that appear in gravitational collapse. There has been a tremendous study on the analysis to find various regular black hole solutions \cite{AyonBeato:1998ub, regular:2005gi, Bronnikov:2000vy, Zaslavskii:2009kp, Lemos:2011dq, Singh:2017bwj}. The regular metric has also been investigated in higher dimensions also to study its thermal and optical properties \cite{Ahmed:2022qge, Singh:2021nvm, Jusufi:2020agr, Amir:2020fpa}. However, the regular metric \cite{regular:2005gi} is also inspired by Bardeen's idea, supported by finite density and pressures, which go off rapidly as the distance increases and are treated as a cosmological constant radius falls off very near to the origin. The formation and evaporation process of a  regular \cite{regular:2005gi} black hole could be explored up to a minimum size $l$. Furthermore, in the regular spacetime \cite{regular:2005gi}, the resulting stress-energy tensor satisfies the weak energy condition but may violate the strong energy condition. The static spherically symmetric regular metric in ($t,r,\theta,\phi$) coordinates read \cite{regular:2005gi}
\begin{eqnarray}\label{metric1}
\mathrm{ds^2}=-\left(1-\frac{2\tilde{m}(r)}{r}\right)dt^2+\frac{dr^2}{\left(1-\frac{2\tilde{m}(r)}{r}\right)}+r^2d\Omega_2^2
\end{eqnarray}
where $d\Omega_2^2=d\theta^2+\sin^2\theta d\phi^2$ and 
\begin{eqnarray}
\tilde{m}(r)=\frac{M r^3}{r^3+g^3},\;r\geq 0.
\label{massf}
\end{eqnarray}
It is an exact solution of general relativity coupled to NED, encompassing the Schwarzschild solution as a particular case ($g=0$). 
An analysis of zeros of $f(r)=0$ reveals a critical value of mass $M=3g/2^{5/3}$ and the radius $r_*=2^{1/3}g$ so that $f(r)$ has a double zeros at $r= r_*$ if $M=M_*$, two zeros at $r=r_\pm$ if $M>M_*$ and no zero if $M<M_*$ \cite{regular:2005gi}. These cases correspond respectively, to an extremal black hole with degenerate horizons, a black hole with Cauchy and event horizons, and no black hole. At large and small $r$ values the regular spacetime with total mass $M$, are having the metric function as 
\begin{eqnarray}
 1-\frac{2M}{r},\;r\to \infty,\;\;\;1-\frac{2M}{g^3}r^2,\;r\to 0.
\end{eqnarray}
 To address the regularity of solution, we  study the behaviour of following curvature invariants:
$\mathrm{R}$, $\mathrm{R_{\mu\nu}R^{\mu\nu}}$, and $\mathrm{R_{\mu\nu\rho\sigma}R^{\mu\nu\rho\sigma}}$
which in the limit $r\to 0$ do not diverge but instead are bounded and finite, indicating that the spacetime (\ref{metric1}) is regular everywhere \cite{Bambi:2013ufa}. A study of thermal behaviours and their evaporation process for the regular black holes has been extensively analyzed \cite{Myung:2007qt}. Thermal fluctuations on the thermodynamic have been investigated \cite{Pourhassan:2016qoz} by modifying the regular black holes. Recently, Fan \cite{fan} invoked the NED to find the exact solution of regular-AdS black holes.\\
However, the issue is that observations cannot test non-rotating black holes, as black hole spin is essential in any astrophysical process and rotating counterparts of these regular metrics, including of the Bardeen and Hayward regular black holes were obtained \cite{Ghosh:2014pba,Toshmatov:2014nya,Amir:2015pja,Ghosh:2014hea}. Also, as per no-hair theorem--the stationary axially symmetric black hole candidates are characterized by their mass ($M$), the rotation parameter ($a$), and the charge ($Q$), but still they lack direct observation and actual nature of them are not yet verified. This opens an area of active research for studying various properties for black holes that are a regular modification to the Kerr black holes. In this direction, the nature of rotating regular black holes can be tested as astrophysical black hole candidates, such as Cygnus $X-1$, using their deviation parameters \cite{Bambi:2011mj, Bambi:2014nta, Bambi:2017yoz, Debnath:2015hea}.

In recent years, the study of black holes thermodynamics has attracted attention due to its many interesting and exciting features, including Hawking radiation, black hole entropy, etc. It may provide a possible way to deepen our understanding of quantum gravity. More specifically, the idea of including the cosmological constant $\Lambda$ in the first law of black hole mechanics provides us with a consistent way to study the phase space thermodynamics of the (anti)-dS spaces. The black holes in dS spaces usually comprise a cosmological horizon besides Cauchy and event horizons. They emit thermal radiation at different temperatures, and the black holes in dS spaces are thermodynamically unstable. As the separate study of the horizons thermodynamics of the dS black holes makes the system thermodynamically unstable, we shall have the relationship between the thermodynamic quantities on the two horizons, which may give rise to an effective way to investigate the thermodynamics quantities in dS spacetime. Based on this logic, many works have been proposed to analyze the critical behaviour of the effective thermodynamic quantities to show that when considering the relation between the two horizons in dS spacetime, there exists a phase transition and critical phenomena similar to the ones in a van der Waals liquid-gas system. The effective thermodynamics of rotating de Sitter spacetime (Kerr-dS and Kerr-Newman-dS) are studied and found to follow the van der Waals type behaviour \cite{Guo:2015waa, Kubiznak:2015bya, Zhang:2014jfa}. 
 Recently a possible analysis of the effective thermodynamics quantities of a regular de Sitter spacetime has been investigated \cite{Ali:2020bgc}. Our paper will investigate a similar phenomenon for the rotating regular-dS black holes and find their thermodynamic quantities. The role of the nonlinear parameter $g$ present in the theory will be emphasized in analytical and numerical ways.\\ 

The paper is organized as follows:  We first review the rotating regular spacetimes in Sec.~\ref{rotregular} and also discuss the thermodynamics. We present an exact rotating regular-dS spacetime in Sec.~\ref{horizons} to study its horizon structure and the effective thermodynamic quantities related to the rotating regular-dS black holes. As a part of the completeness of the study of the thermal properties, we present the corrected entropy for the de Sitter black holes in Subsec.~\ref{correntropy_for}. The paper ends with concluding remarks in Sec.~\ref{conclusion}.
 
\section{Rotating regular black holes}\label{rotregular}
Next, we discuss the thermodynamics of rotating regular black holes in the presence of a nonlinear electrodynamic field. Before going into the main discussion, we write the gauge potential for a magnetically charged black hole. The electromagnetic potential for the magnetically charged rotating black hole was recently studied and derived using the usual Newman-Janis procedure, which leads to having the form \cite{Toshmatov:2017zpr, Erbin:2016lzq}
\begin{eqnarray}
\label{rot_gauge}
A_\mu=-\frac{g\;a\;\cos\theta}{\Sigma}\delta_\mu^t+\frac{g(r^2+a^2)\cos\theta}{\Sigma}\delta_\mu^\phi.
\end{eqnarray}
The guage potential in the absence of the rotation ($a=0$), reduces to the expression for spherically symmetric spacetime is writen as $A_\mu=g\cos\theta\delta_\mu^\phi$.
With this guage potential in hand, we have Maxwell field invariant to be \cite{Toshmatov:2017zpr}
\begin{eqnarray}
\label{lagran_rot}
\mathcal{F}=\frac{2g^2\left(\left(\Sigma-2r^2\right)^2-4a^4\cos^4\theta\right)}{\Sigma^4}.
\end{eqnarray}
The Einstein's field equations lead to have the expressions of Lagrangian  \cite{Toshmatov:2017zpr} in the form
\begin{eqnarray}
\label{lag_rot}
\mathcal{L}(r)&=&\frac{4r^2\left(2r^2\Sigma\;m^{\prime\prime}+5\;m^\prime\left(\left(\Sigma-2r^2\right)^2-8r^2\Sigma+4r^2\right)\right)}{\Sigma^4},
\end{eqnarray}
where $^{\prime}$ and $^{\prime\prime}$ denotes, respectively, the first and second derivatives with respect to the radial coordinate $r$.
The rotating regular black holes are described by three parameters, e.g., the mass $M$, the rotation parameter $a$, the nonlinear parameter $g$. The rotating regular metric belongs to prototype family of non-Kerr solutions, which in Boyer-Lindquist coordinates has same form as that of Kerr metric with mass $m$ replaced by some mass function $\tilde{m}$ which contains an additional deviation parameter $g$  from the NED, which in Boyer-Lindquist coordinates $\left(t,r,\theta,\phi\right)$ coordinate reads \cite{Amir:2015pja}
\begin{eqnarray}\label{metric}
ds^2&=&-\left(1-\frac{2\tilde{m}(r)r}{\Sigma}\right)dt^2-\frac{4a\tilde{m}(r)r\sin^2\theta}{\Sigma}dtd\phi+\frac{\Sigma}{\Delta}dr^2+\Sigma d\theta^2\nonumber\\
&&\sin^2\theta\left(r^2+a^2+\frac{2a^2\tilde{m}(r)r\sin^2\theta}{\Sigma}\right)d\phi^2,
\end{eqnarray}
where $\tilde{m}(r)$ is defined in Eq.~(\ref{massf}), and $\Delta=r^2+a^2-2\tilde{m}(r)r,\;\Sigma=r^2+a^2\cos^2\theta,$ $a$ is the black hole spin parameter and $g$ is the magnetic charge arising from the NED, which measures the potential deviation from the Kerr metric ($g=0$). The rotating regular  black hole metric (\ref{metric}) is independent of $t,\,\,\phi$,  which implies existence of  two Killing vectors given by $\eta^\mu=\delta{^\mu_t}$ and $\xi^\mu=\delta{^\mu_\phi}$. 
The horizons of rotating regular black holes are solution of  \cite{Amir:2015pja}
\begin{eqnarray}\label{horizon}
&&\left(\eta^\nu,\xi_\mu\right)^2-(\eta^\mu\eta_\mu)(\xi^\mu\xi_\mu)=g{_{t\phi}}^2-g_{tt} g_{\phi\phi}=\Delta=0,
\end{eqnarray}
It turns out that for a given value of the parameter $a$, there exists an extremal value $g_E$ for which Eq.~(\ref{horizon}) admits double root corresponding to the extremal black holes, while for $g<g_E$, the equation (\ref{horizon}) can admit two possible roots ($r_\pm$) corresponding to inner (Cauchy) ($r_-$) and outer (Event) ($r_+$) horizons  \cite{Amir:2015pja}. 
\subsection{Thermodynamics}\label{thermodynamics}
Next, we discuss  thermodynamics of the rotating regular black holes on the black hole event horizon. The mass of  the rotating regular black holes  in terms of the event horizon radius $r_+$ is obtained by solving $\Delta(r_+)=0$, which yields
\begin{eqnarray}\label{massh}
M_+=\frac{\left(r_+^2+a^2\right)\left(r_+^3+g^3\right)}{2r_+^4}.
\end{eqnarray}
Obviously, $g=0$ leads to the Kerr black hole mass \cite{Czinner:2017tjq}.
The Killing field $\chi{^\mu}$ associated with the Killing vectors is expressed as $\chi{^\mu}=\eta^{\mu}+\Omega\xi^{\mu}$ \cite{Modesto:2010rv}. Here $\Omega$ is the angular velocity at the black hole event horizon which can be obtained with the requirement that $\chi{^\mu}$ is a null generator of the event horizon of the black holes i.e. $\chi{^\mu}\chi{_\mu}=0$, which leads to 
\begin{eqnarray}
\label{Omega1}
g_{tt}+2\Omega g_{t\phi}+\Omega^2 g_{\phi\phi}=0.
\end{eqnarray}
The $\Omega$ is obtained from Eqs.~(\ref{metric}) and (\ref{Omega1}) as
\begin{eqnarray}
\Omega=\omega\pm\frac{\Sigma\Delta^{\frac{1}{2}}}{\sin\theta\left[(a^2\sin^2\theta+\Sigma)^2-\Delta a^2\sin^2\theta\right]}.
\end{eqnarray}
On the event horizon,  $\Delta(r_+)=0$, thereby the angular velocity reduces to
\begin{eqnarray}\label{omega}
\Omega_+=\omega\vert_{r=r_+}=\frac{a}{r_+^2+a^2}.
\end{eqnarray}
The angular momentum of the black hole at the event horizon is derived as
\begin{eqnarray}
\label{angular_momentum}
J_+=aM_+=\frac{a\left(r_+^2+a^2\right)\left(r_+^3+g^3\right)}{2r_+^4}.
\end{eqnarray}
The surface gravity $\kappa=\sqrt{-\frac{1}{2}\nabla_{\mu}\chi{_\nu}\nabla{^\mu}\chi{^\nu}}$ \cite{Modesto:2010rv} of rotating regular black holes at the event horizon yield, we have the Hawking temperature of the black hole, via $T_+=\kappa/2\pi$ which for the rotating regular black hole metric (\ref{metric}) is given by
\begin{eqnarray}
\label{htemp}
T_+=\frac{3r_+^3\left(r_+^2+a^2\right)+2r_+^2\left(r_+^3+g^3\right)-4\left(r_+^2+a^2\right)\left(r_+^3+g^3\right)}{4\pi r_+\left(r_+^2+a^2\right)\left(r_+^3+g^3\right)}.
\end{eqnarray}
\begin{figure}
\begin{tabular}{c c c c}
 \includegraphics[scale=0.62]{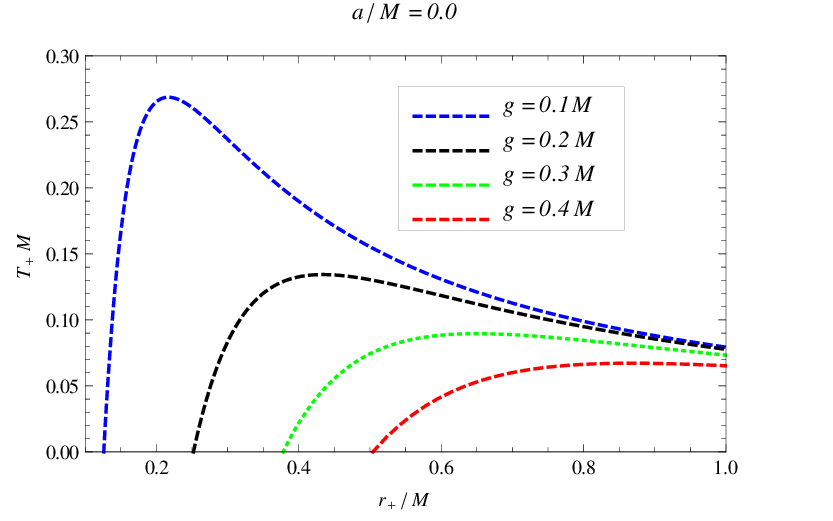}
&\includegraphics[scale=0.62]{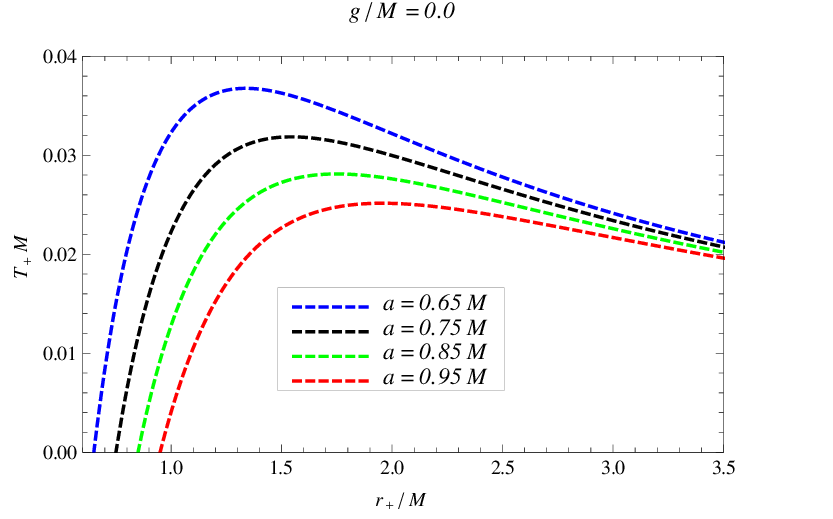}\\
\includegraphics[scale=0.62]{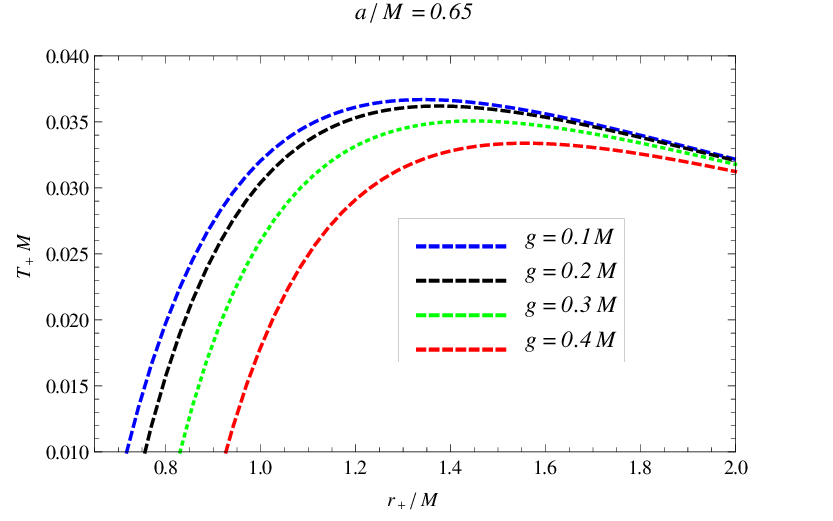}
&\includegraphics[scale=0.62]{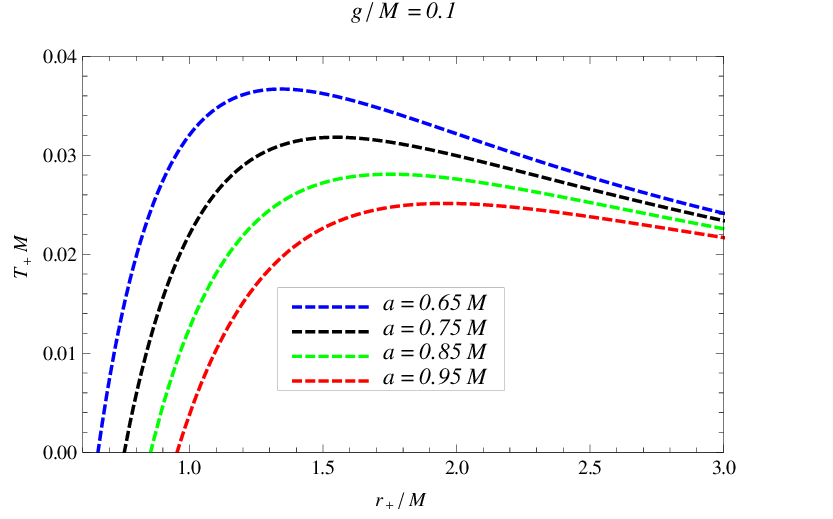}
\end{tabular}
\caption{The plot showing the behaviour of Hawking temperature $T_+M$ vs horizon radius $r_+/M$ for rotating regular black holes and a comparison with the temperature of spherical regular black holes ($a=0$) and the Kerr black holes ($g=0$).}\label{figtemph2}
\end{figure}
The behaviour of the temperature as a function of the event horizon radius is depicted in Fig.~\ref{figtemph2} for different values of the charge $g$, and the rotation parameter $a$. The figure suggests that the rotating regular black holes ($g\neq 0$) are colder than the Kerr black holes ($g=0$). The Schwarzshild black hole temperature, i.e., $T_+=1/4\pi r_+ $ shows the divergent behaviour as $r_+\to 0$. When $g=0$, Eq.~(\ref{htemp})  gives Kerr black holes temperature \cite{Czinner:2017tjq}. 

The temperature (\ref{htemp}) of the rotating regular black holes, like the Kerr black holes, does not diverge but shows a finite peak at short distances comparable to the Planck scale \cite{Myung:2007qt}. The peak shifts to the right and smaller values with increasing the charge $g$. Fig.~\ref{figtemph2} shows the temperature of the rotating regular and Kerr black holes  grows to a maximum $T_+^C$, at $r_+=r_+^C$ and $M_+=M_+^C$ and then drops down to zero at $r_+=r_+^*$ and $M_+=M_+^*$, corresponding to the extremal temperature $T_+^*=0$. Consequently, the evaporation process is split into two important branches, the right branch $r_+^C<r_+<\infty$ called the early stage of the evaporation process and the left branch $r_+^*<r_+<r_+^C$ called the quantum cooling evaporation process \cite{Myung:2007qt}.
\begin{table}[h]
    \begin{center}
        \begin{tabular}[b]{|c|c|c|c|c|c|c|c|c|c|c|c|c|c|c|c|c|}
            \hline 
            &\multicolumn{3}{c}{$a=0.65$} &&\multicolumn{4}{c|}{$a=0.85$}
            &\multicolumn{4}{c|}{$a=0.95$}
            \\ \hline &$g=0.1$& 
            $g=0.2$& $g=0.3$& $g=0.4$&
            $g=0.1$&
            $g=0.2$& $g=0.3$& $g=0.4$
            &
            $g=0.1$&
            $g=0.2$& $g=0.3$& $g=0.4$
            \\ \hline & & & & & & & & & & & & 
            \\
            $r_+^{C}$
            &1.3426& 1.3742& 1.4469& 1.5569
             &1.7522& 1.7714& 1.8193& 1.8994 
             &1.9575& 1.9730& 2.0125&2.08096
            \\ \hline & & & & & & & & & & & & 
            \\
            $M_+^{C}$&0.8289& 0.8434&0.8772&0.9296
                   &1.0825& 1.0912&1.1132& 1.1505 
                   &1.2094&1.2165& 1.2345&1.2662
            \\ \hline & & & & & & & & & & & & 
            \\
            $T_+^{C}$&0.0366&0.0361&0.0351&0.0334
 &0.0281&0.0279&0.0275&0.0266
 &0.0251&0.025&0.0247&0.0242
            \\ \hline 
        \end{tabular}
        \caption{The tabulated values of the critical radius $r_+^C$, the critical mass $M_+^{C}$ and the critical temperature $T_+^{C}$ for different values of the charge parameter $g$ and rotation parameter $a$.\label{rem1}}
    \end{center}
\end{table}
\begin{table}[h]
    \begin{center}
        \begin{tabular}[b]{|c|c|c|c|c|c|c|c|c|c|c|c|c|c|c|c|c|}
            \hline 
            &\multicolumn{3}{c}{$a=0.65$} &&\multicolumn{4}{c|}{$a=0.85$}
            &\multicolumn{4}{c|}{$a=0.95$}
            \\ \hline &$g=0.1$& 
            $g=0.2$& $g=0.3$& $g=0.4$&
            $g=0.1$&
            $g=0.2$& $g=0.3$& $g=0.4$
            &
            $g=0.1$&
            $g=0.2$& $g=0.3$& $g=0.4$
            \\ \hline & & & & & & & & & & & & 
            \\
            $r_+^*$
            &0.6568&0.6967&0.7698&0.8607&0.8540&0.8801&0.9356&1.0132
            &0.9532& 0.9747&1.0229&1.0936
            \\ \hline & & & & & & & & & & & & 
            \\
            $M_+^*$&0.6523&0.6669&0.6983&0.7436&0.8513&0.8604&0.8820
            &0.9162&0.9511&0.9585&0.9766&1.0063
            \\ \hline 
        \end{tabular}
        \caption{The tabulated values of the extremal horizon $r_+^*$ and the extremal mass $M_+^*$ for different values of the charge parameter $g$ and rotation parameter $a$.\label{rem2}}
    \end{center}
\end{table}
The first law for the rotating regular black holes, considering the magnetic  charge $g$ as a variable quantity conjugate to the corresponding magnetic potential can be written as
\begin{eqnarray}
\label{firstlawh}
dM_+=T_+dS_++\Omega_+ dJ+\Psi_+dg,
\end{eqnarray}
where $\Psi_+$ is the potential conjugate to charge $g$. The temperature, angular velocity and the magnetic potential,respectively,  can be obtained from
\begin{eqnarray}
T_+&=&\left(\frac{\partial M_+}{\partial S_+}\right)_{J_+,g},\;\Omega_+ =\left(\frac{\partial M_+}{\partial J_+}\right)_{S_+,g},\;\Psi_+ =\left(\frac{\partial M_+}{\partial g}\right)_{S_+,J_+}.
\end{eqnarray}
The entropy of the rotating regular black holes, using Eqs.~(\ref{firstlawh}) and (\ref{massh}), for constant $g$, yield
\begin{eqnarray}\label{entropy}
S_+=\pi\left[r_+^2-\frac{2g^3}{r_+}\left(1+\frac{a^2}{3r_+^2}\right)\right]+C_0,
\end{eqnarray}
where, $C_0$ is an integration constant and chosen to be $C_0=\pi{a^2}$. Therefore the entropy of the rotating regular black hole is written as $S_+=S_{\text{Kerr}}+S_{\text{corr}}$, 
where $S_{\text{Kerr}}=\pi(r_+^2+a^2)$ and $S_{\text{corr}}=-\frac{2\pi g^3}{r_+}\left(1+\frac{a^2}{3r_+^2}\right)$. Since there is a relative ``$-$" sign indicating that the regular black hole is less disordered than the singular counterpart like Kerr black holes. Obviously, the entropy of rotating regular black holes no longer obeys the Bekenstein area law. In the limit $a\to 0$, Eq.~(\ref{entropy}) reduces to the entropy of the spherical regular black holes \cite{Banerjee:2012,J Man:2014}. Further, when $a=g=0$, we have $S_+=\pi r_+^2$, which is the entropy of the Schwarzschild black hole and also can be rewritten as the Bekenstein's area law, i.e., $S_+=A/4$, where $A$ is the area of the event horizon. \\
The heat capacity is related to the local thermal stability and the black holes are unstable to thermal radiation if it has negative heat capacity  and for positive heat heat capacity the black holes are stable to thermal radiation. The heat capacity of Schwarzschild black hole is always negative which continuously losing its energy through Hawking radiation \cite{Gross:1982cv} thereby it is thermodynamically unstable. For the spherically symmetric regular black holes the heat capacity has a phase transition which has been discussed \cite{Myung:2007qt, Myung:2007av} at some horizon radius such that there exists a discontinuity in the heat capacity. The heat capacity of a black hole is defined by
\begin{eqnarray}
C_+=\frac{\partial M_+}{\partial T_+}=\frac{\partial M_+}{\partial r_+}\frac{\partial r_+}{\partial T_+},
\end{eqnarray}
Using the Eqs. (\ref{massh}) and (\ref{htemp}), the heat capacity for rotating regular regular black hole reads
\begin{eqnarray}\label{heatcap}
C_+=-\frac{2\pi\left[\left(r_+^2+a^2\right)^2\left(r_+^3+g^3\right)^2\left(r_+^5-\left(a^2r_++2g^3\right)r_+^2-4g^3a^2\right)\right]}{r_+^3\left[r_+^{10}-\left(4a^2r_+^2+a^4+10g^3\right)r_+^6-2g^3\left(16a^2r_++g^3\right)r_+^4-2g^3a^2\left(7a^2r_++5g^3\right)r_+^2-4a^4g^6\right]},
\end{eqnarray}
\begin{figure}
    \begin{tabular}{c c c c}
        \includegraphics[scale=0.62]{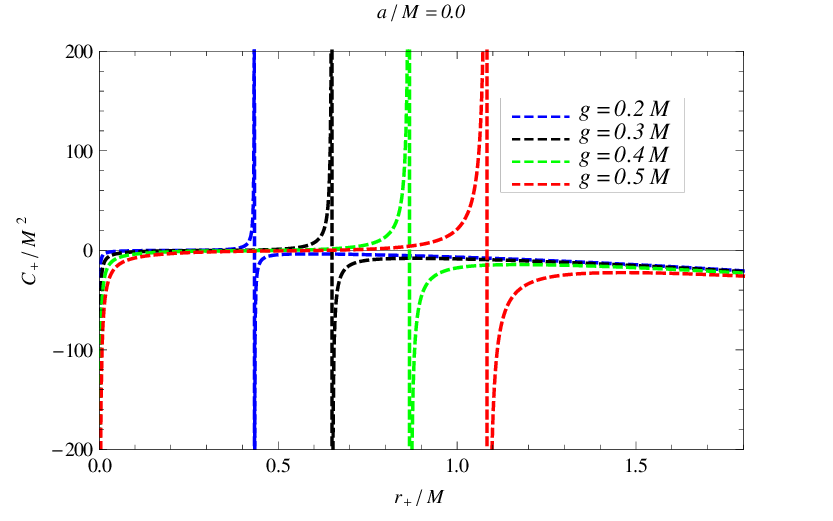}\hspace{-0.2cm}
        &\includegraphics[scale=0.62]{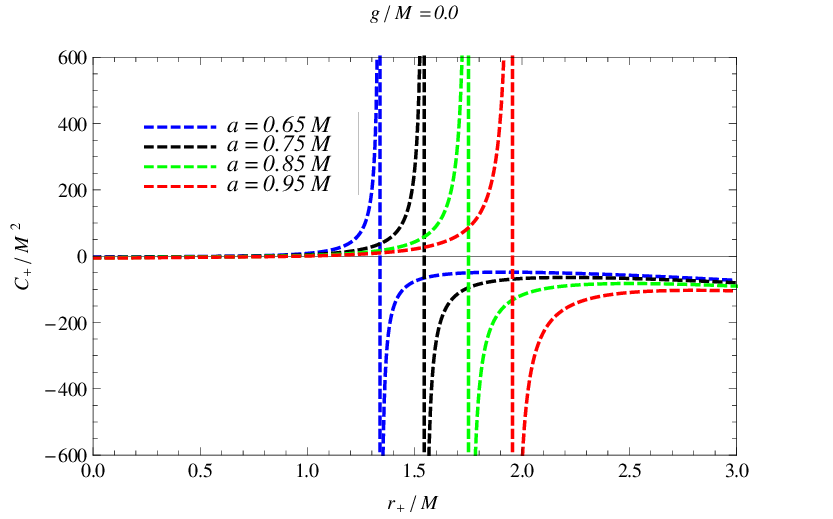}\\
        \includegraphics[scale=0.62]{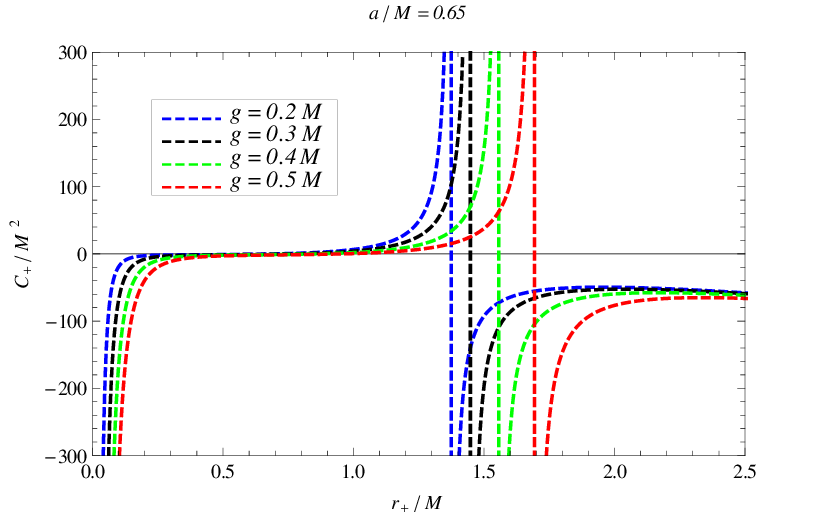}\hspace{-0.2cm}
        &\includegraphics[scale=0.62]{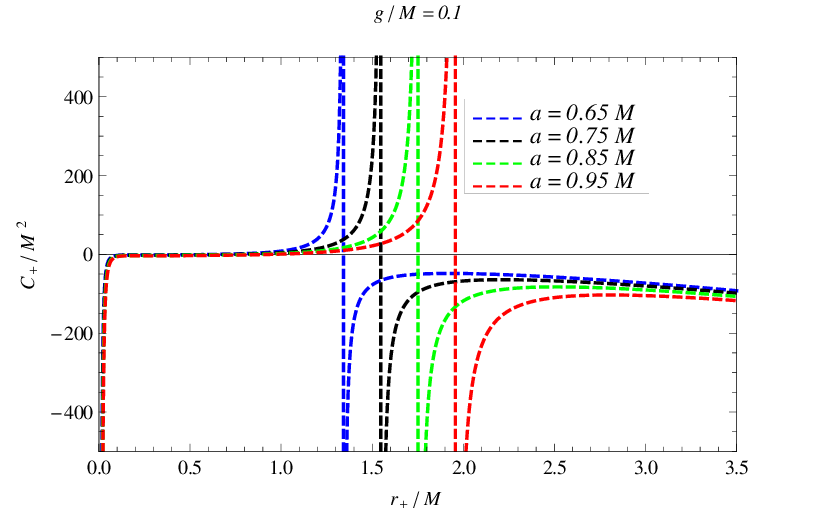}
    \end{tabular}
    \caption{The plot showing the behaviour of heat capacity $C_+/M^2$ vs horizon radius $r_+/M$ for rotating regular black holes and a comparison with the corresponding heat capacity of  spherical regular black holes ($a=0$) and Kerr black holes ($g=0$).}\label{figheatcaph2}
\end{figure}
Fig.~\ref{figheatcaph2} shows that the rotating regular black holes is stable in the region $r_+^*<r_+<r_+^C$, where $r_+^C$ is the critical horizon radius where the heat capacity diverges, and the corresponding temperature becomes maximum  while the black holes is thermodynamically unstable in the range $r_+^C<r_+<\infty$. When $a=g=0$, the heat capacity (\ref{heatcap}) becomes $C_+=-2\pi r_+^2$, that of the Schwarzschild black hole. The heat capacity for $g=0$, reduces to the expression of Kerr black hole. When $r_+=r_+^*$, both the heat capacity and temperature approach to zero. We note that the temperature of the black holes vanishes at $r_+^*$, where $r_+^*$ corresponding to the extremal value of the black hole horizons with extremal mass $M^*_+$ of the rotating regular black holes. When $r_+<r_+^*$, the temperature corresponds to the negative value and is not physically significant. The minimum mass corresponding to the extremal horizon radius $r_+^*$ is known as the remnant mass. The size and the mass of the remnant increase with the increase in the charge parameter $g$ and the rotation parameter $a$ (see Table.~\ref{rem1}).

\section{The rotating regular-dS black holes and effective thermodynamics}\label{horizons}
\begin{figure}
\begin{tabular}{c c c c}
\includegraphics[scale=0.62]{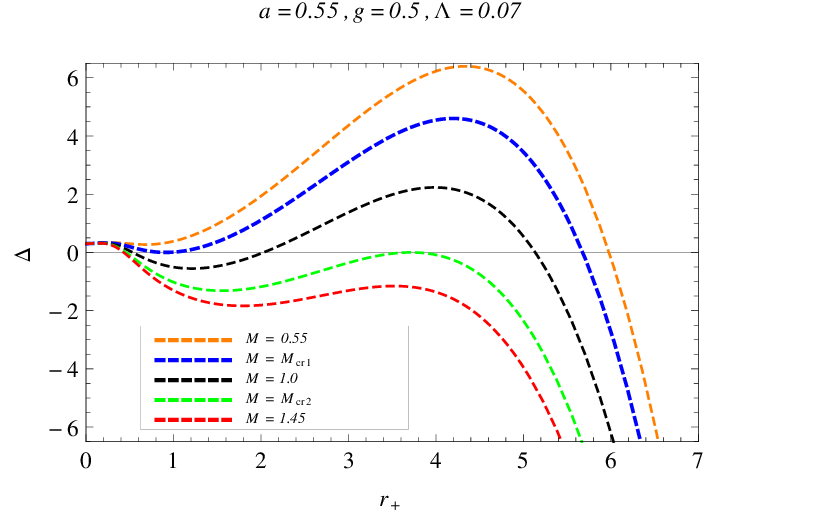}\hspace{-0.2cm}
&\includegraphics[scale=0.62]{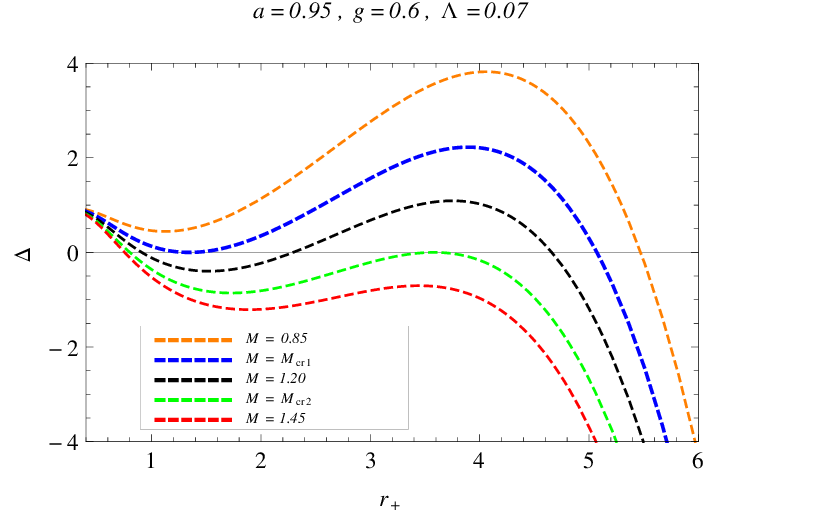}\\
\includegraphics[scale=0.62]{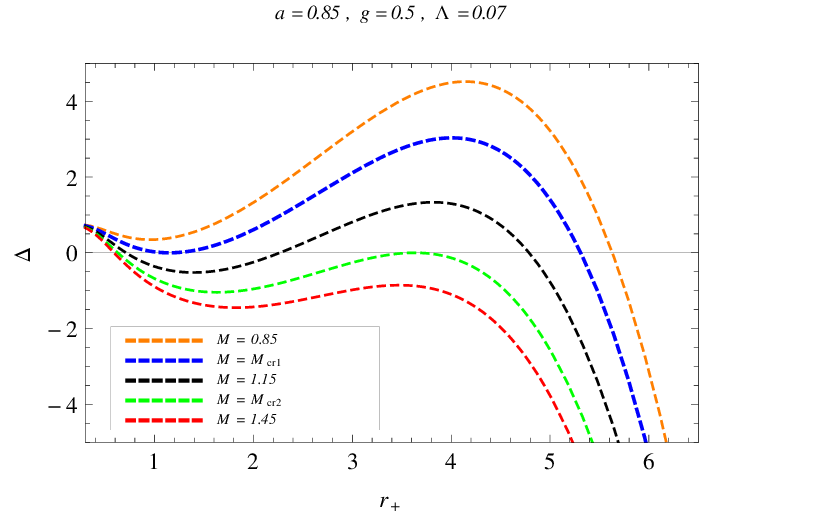}\hspace{-0.2cm}
&\includegraphics[scale=0.62]{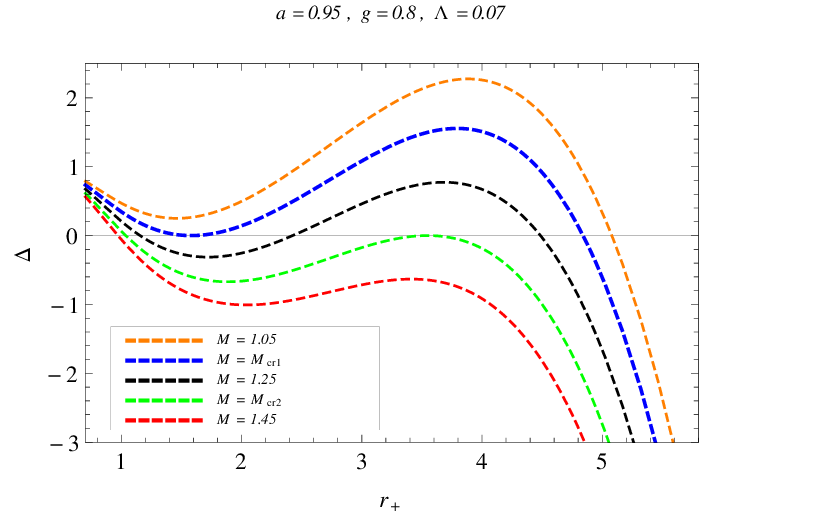}\\
\end{tabular}
\caption{The plot of $\Delta_r$ vs $r$ for rotating regular-dS black hole for different values of the parameters $a$ and $g$ at a fixed $\Lambda$.}\label{HDS}
\end{figure}

In this section, we extend the rotating regular regular black holes in the asymptotically dS spacetimes whose metric in the Boyer-Lindquist coordinates $\left(t,r,\theta,\phi\right)$ reads
\begin{eqnarray}\label{regrotds}
ds^2=-\frac{\Delta_r}{\Sigma}\left(dt-\frac{a\sin^2\theta}{\Xi}d\phi\right)^2+\frac{\Sigma}{\Delta_r}dr^2+\frac{\Sigma}{\Delta_{\theta}}d\theta^2+\frac{\Delta_{\theta}\sin^2\theta}{\Sigma}\left(adt-\frac{r^2+a^2}{\Xi}d\phi\right)^2,
\end{eqnarray}
where
\begin{eqnarray}\label{horz}
\Sigma&=& r^2+a^2\cos^2\theta,\,\,\ \Xi=1+\frac{\Lambda}{3} a^2,\\
\Delta_r&=& \left(r^2+a^2\right)\left(1-\frac{\Lambda}{3} r^2\right)-2\tilde{m}(r)r,\,\,\,\ \Delta_{\theta}=1+\frac{\Lambda}{3} a^2\cos^2\theta,
\end{eqnarray}
where $\tilde{m}(r)$ is the mass function for rotating regular-dS black holes defined in Eq.~(\ref{massh}) and $\Lambda$ is the positive cosmological constant.  An analysis of the horizons of the rotating regular black holes has been extensively discussed \cite{Ghosh:2015pra} and shown that there exists  values of the parameters $a\;\text{and}\;g$ such that rotating regular black holes admit two horizons, viz, the inner Cauchy and outer event horizons. Here, we extend  a discussion the horizon structures of the rotating regular-dS black holes and addition of the $\Lambda$ makes a significant contribution. Indeed, a positive cosmological constant $\Lambda$ for different values of the mass parameter $M$, the nonlinear parameter $g$, and the rotation parameter $a$, we get three horizons (c.f.~Fig.~\ref{HDS}) viz., the inner Cauchy horizon $r_-$, the outer event horizon $r_+$, the outermost cosmological horizon $r_C$, i.e., $r_-<r_+<r_C$. There exist two extreme conditions corresponding to two critical masses $M_{\text{cr1}}\;\text{and}\;M_{\text{cr2}}$ such that  when $M_{\text{cr1}}<M<M_{\text{cr2}}$, we have three distinct horizons, while $M=M_{\text{cr1}}\;\text{and}\;M=M_{\text{cr2}}$, correspond, respectively, to the degenerate Cauchy and event horizons when $r_-=r_+=r_+^E$ and the degenerate event and cosmological horizons when $r_+=r_C=r_C^E$. On the other hand, when $M<M_{\text{cr1}}$, there exists cosmological horizon only, and hence it puts a lower bound on the mass parameter of the black holes \cite{Dymnikova:2010zz}. This minimal configuration therefore corresponds to the mass of a remnant \cite{Dymnikova:2010zz}. The configuration with the critical mass $M_{\text{cr2}}$ corresponds to a regular modification to the Narai-type solution. In the following we study the thermodynamic properties related to the black hole event horizon and the cosmological horizon and find the effective thermodynamic quantities relating the two horizons.
Next, we derive the thermodynamic quantities associated with rotating regular-dS black holes in terms of the event horizon and cosmological horizons. The black hole event horizon and cosmological horizon, respectively, satisfy the equations $\Delta_r(r_+)=0$ and $\Delta_r(r_C)=0$ \cite{Guo:2015waa, Kubiznak:2015bya, Zhang:2014jfa}. By solving them simultaneously, we obtain
\begin{eqnarray}
\label{thermoeff2}
2m&=&\frac{\left(r_+^2+a^2\right) \left(r_C^2+a^2\right) \left(r_+^3+g^3\right) \left(r_C^3+g^3\right) \left(r_++r_C\right)}{2 r_+^2 r_C^2 \left(r_+^2 r_C^2 \left(r_+^2+r_++r_C^2\right)-a^2 \left(g^3 (r_++r_C)-r_+^2 r_C^2\right)\right)},
\end{eqnarray}
\begin{eqnarray}
\label{thermoeff3}
\Xi&=&\frac{r_+^3r_C^3\left[r_+r_C\left(r_+^2+r_+r_C+r_C^2+2a^2\right)-a^4\right]-a^2g^3\left(r_++r_C\right)\left(a^2(r_+^2+r_C^2)+2r_+^2r_C^2\right)}{r_+^2r_C^2\left[r_+^2r_C^2\left(r_+^2+r_+r_C+r_C^2+a^2\right)-a^2g^3(r_++r_C)-r_C^3\left(r_++r_C\right)\right]},
\end{eqnarray}
\begin{eqnarray}
\label{thermoeff4}
\frac{\Lambda}{3}&=&\frac{r_+^3r_C^3\left(r_+r_C-a^2\right)-g^3 (r_++r_C) \left(r_+^2r_C^2+a^2 \left(r_+^2+r_C^2\right)\right)}{r_+^2r_C^2\left[r_+^2r_C^2\left(r_+^2+r_+r_C+r_C^2+a^2\right)-a^2g^3\left(r_++r_C\right)\right]}
\end{eqnarray}
which for $g=0$ and $a=0$, respectively, reduce to the expressions of the Kerr-dS black holes \cite{Guo:2015waa} and regular-dS black holes \cite{Nam:2018tpf}. The Hawking temperatures of the event horizon and cosmological horizon, respectively, are given by $T_+=\kappa_+/2\pi$, $T_C=-\kappa_C/2\pi$, where $\kappa_+$ and $-\kappa_C$ are surface gravities of the two horizons with $\kappa_C>0$ and $\kappa_+\geq \kappa_C$. The Hawking temperatures for the rotating regular dS black holes are given by
\begin{eqnarray}
\label{therEV}
T_{+,C}&=&\frac{r_{+,C}\Bigg[\left(r_{+,C}^6+2g^3r_{+,C}^3-m\left(4g^3r_{+,C}^2+r_{+,C}^5\right)\right)-\left(2r_{+,C}^2+a^2\right)\left(r_{+,C}^3+g^3\right)\frac{\Lambda}{3}\Bigg]}{2\pi\left(r_{+,C}^2+a^2\right)\left(r_{+,C}^3+g^3\right)},
\end{eqnarray}
where $m$ and $\Lambda$ are, respectively, already defined in the Eqs.~(\ref{thermoeff2}) and (\ref{thermoeff4}).
 Eq.~(\ref{therEV}), when $g=0$, reduces to that of the temperature expressions of Kerr-dS black holes \cite{Guo:2015waa}, and  for $a=0$, one can get the temperatures of spherical regular-dS black holes \cite{Nam:2018tpf}.
The Bekestein-Hawking entropy associated with the two horizons are found in \cite{Guo:2015waa, Kubiznak:2015bya}. The entropies of the black hole horizon and the cosmological horizon are different but the mass and angular momentum, have the same values irrespective of the horizons but with the opposite sign \cite{Sekiwa:2006qj}. The angular momentum and the angular velocity at the asymptotic infinity read
\begin{eqnarray}\label{angularM}
J&=&\frac{a\;m}{\Xi^2}=\frac{a}{2\Xi^2}\frac{\left(r_+^2+a^2\right) \left(r_C^2+a^2\right) \left(r_+^3+g^3\right) \left(r_C^3+g^3\right) \left(r_++r_C\right)}{r_+^2 r_C^2 \left(r_+^2 r_C^2 \left(r_+^2+r_++r_C^2\right)-a^2 \left(g^3 (r_++r_C)-r_+^2 r_C^2\right)\right)},\;\Omega_{\infty}=\frac{\Lambda}{3}a.
\end{eqnarray}
The thermodynamics quantities of the event and cosmological horizon are related, respectively, via the first law black hole thermodynamics  \cite{Guo:2015waa, Kubiznak:2015bya, Zhang:2014jfa}
\begin{eqnarray}\label{firstlaw}
\delta M&=&T_+\delta S_++\Omega_+\delta J+V_+\delta P,\nonumber\\
\delta M&=&-T_C\delta S_C+\Omega_C\delta J+V_+\delta P,
\end{eqnarray}
where
\begin{eqnarray}
\Omega_+&=&=\frac{a r_+^2\left(g+r_C\right)\left(r_++r_C\right)\left(a^2+r_C^2\right)\left(g^2-gr_C+r_C^2\right)}{r_C^2\left(r_+^2+a^2\right)\left[r_+^2r_C^2\left(r_+^2+r_+r_C+r_C^2+a^2\right)-a^2g^3\left(r_++r_C\right)\right]},\;\Omega_C=\Omega_+(r_+\Longleftrightarrow r_C)\nonumber\\
\end{eqnarray}
where we have inserted $\Lambda$ using from Eq.(\ref{thermoeff4}).
The thermodynamics volumes relating the black hole horizon and cosmological horizons, respectively, read
\begin{eqnarray}\label{vols}
V_{+,C}&=&\frac{2\pi(r_{+,C}^2+a^2)}{3r{+,C}\Xi}\left(2r_{+,C}^2+a^2+\frac{\Lambda a}{3}a^2r_{+,C}^2\right),
\end{eqnarray}
where  the cosmological constant $\Lambda$ is related to the conjugate pressure via
$\Lambda=-8\pi P.$
If $a\ll r_+$ \cite{Guo:2015waa}, we have
\begin{eqnarray}\label{VV}
V=V_C-V_+\approx\frac{4\pi}{3}\left(r_C^3-r_+^3\right)
\end{eqnarray}
For a fixed $a$, and from Eq.~(\ref{regrotds}) we obtain
\begin{eqnarray}\label{kappa12}
dr_+\approx \frac{r_+dM}{\left(r_+^2+g^2\right)\kappa_+}+\frac{r_+^2d\Lambda}{6\kappa_+},\,\,\,\,dr_C=\frac{r_CdM}{\left(r_C^2+g^2\right)\kappa_C}+\frac{r_C^2d\Lambda}{6\kappa_C},
\end{eqnarray}
Substituting Eq.~(\ref{kappa12}) into Eq.~(\ref{VV}), we get
\begin{eqnarray}\label{volmef}
dV\approx 4\pi\left(\frac{r_C}{\kappa_C}-\frac{r_+}{\kappa_+}\right)dM+\frac{2\pi}{3}\left(\frac{r_C^4}{\kappa_C}-\frac{r_+^4}{\kappa_+}\right)d\Lambda.
\end{eqnarray}
Obviously, in the absence of magnetic charge ($g$), one can recovers the corresponding expressions for the Kerr dS \cite{Guo:2015waa}. 
\subsection{Thermodynamic quantities}
In the previous section, we have shown that the temperatures associated with event and cosmological horizons may not be equal and thus, the rotating regular-dS system cannot acquire thermal equilibrium. However, the three variables \textit{viz.} $M,\,a,\,\text{and}\,\,\Lambda$ connecting  event and cosmological horizons are interlinked with their Hawking temperatures. Hence, it is expected that thermodynamic quantities at each of the two horizons are related and for this when we discuss the thermodynamical aspects of rotating regular-dS spacetime, we incorporate, in an effective way, the thermodynamics of the two horizons. Interestingly, the special cases for the rotating regular-dS black hole viz, the Nariai black hole and the lukewarm black holes \cite{Kubiznak:2015bya, Dolan:2013ft, Hajian:2016kxx, McInerney:2015xwa, Bhattacharya:2013tq, Pappas:2016ovo} where temperatures are same at  the two horizons. When the  event  and cosmological event horizons coincide apparently, they have the same temperature and correspond to the Narai solutions \cite{Bousso:1996au}. A study of the dS spacetime reveals the rate of emitting particles, from the two horizons, having the energy $\omega$ such that 
\begin{eqnarray}
\Gamma=\text{exp}\left(\Delta S_C\right)\text{exp}\left(\Delta S_+\right)=\text{exp}\left(\Delta S_C+\Delta S_+\right)
\end{eqnarray} 
where $\Delta S_+\,\,\,\,\text{and}\,\,\,\Delta S_C$ are the difference in the entropies, respectively, corresponding to event and  cosmological horizons  when dS black holes emitted particles with energy $\omega$. Thus, the rate of emission of the radiating particles can be thought of as the product of the rate of particles emitted from the black hole horizon and the cosmological horizon. Therefore the cumulative effect of the entropies of the event and cosmological horizon contribute to the effective entropy of the dS spacetime as 
\begin{eqnarray}\label{entropylaw}
S=S_++S_C.
\end{eqnarray}
Utilizing Eq.~(\ref{firstlaw}) into Eq.~(\ref{entropylaw}), we have
\begin{eqnarray}\label{entropyf}
dS=2\pi\left(\frac{1}{\kappa_+}+\frac{1}{\kappa_C}\right)dM-2\pi\left(\frac{\Omega_+}{\kappa_+}+\frac{\Omega_C}{\kappa_C}\right)dJ-2\pi\left(\frac{\Psi_+}{\kappa_+}+\frac{\Psi_C}{\kappa_C}\right)dg+\frac{1}{4}\left(\frac{V_+}{\kappa_+}+\frac{V_C}{\kappa_C}\right)d\Lambda,
\end{eqnarray}
 Utilizing Eqs.~(\ref{volmef}) and (\ref{entropyf}) we can obtain
\begin{eqnarray}
dM=T_{eff}dS+P_{eff}dV+\Psi_{eff}dg+\Omega_{eff}dJ,
\end{eqnarray}
and the corresponding Smarr relation \cite{Cheng:2016bpx, law1st, dolan, mann, Mo:2014mba, Dolan:2011xt, Gunasekaran:2012dq} reads
\begin{eqnarray}
\label{smarreff}
M=2T_{eff}S+\Phi_{eff} g-2P_{eff}V+\Omega_{eff}J.
\end{eqnarray}
We obtain from Eq.(\ref{smarreff}), the thermodynamic quantities of the rotating globally regular-dS black hole,
\begin{eqnarray}\label{teff}
T_{eff}
&\approx &
\frac{\left(\kappa_+-x^4\kappa_C \right)}{2\pi\left(1-x^2\right)\left(1+x+x^2\right)}
\end{eqnarray}
\begin{eqnarray}
\Omega_{eff}
&\approx& 
\frac{\left(\kappa_+-x^4\kappa_C \right)\left(\Omega_+\kappa_C+\Omega_c\kappa_+\right)}{2\pi\left(1-x^2\right)\left(1+x+x^2\right)\kappa_+\kappa_C}
\end{eqnarray}
\begin{eqnarray}\label{peff}
P_{eff}
&\approx&\frac{1}{4\pi r_C}\left[\frac{\left(x^3\kappa_C+\kappa_+\right)}{\left(1-x^2\right)\left(1+x+x^2\right)}\right]
\end{eqnarray}
The limiting case, $x=1$, i.e., when $r_+=r_C$, we have $T_+=T_C=0$ but their surface areas are not zero indicating that they have nonzero entropy  $S$=$S_++S_C$=$2S_+$=$2S_C$. Further, the effective pressure $P_{eff}$ of the rotating regular dS spacetimes approaches to zero but the temperature approaches the steady state value when $x \rightarrow 1$. \\
\begin{figure}
\begin{tabular}{c c c c c c c}
 \includegraphics[scale=0.62]{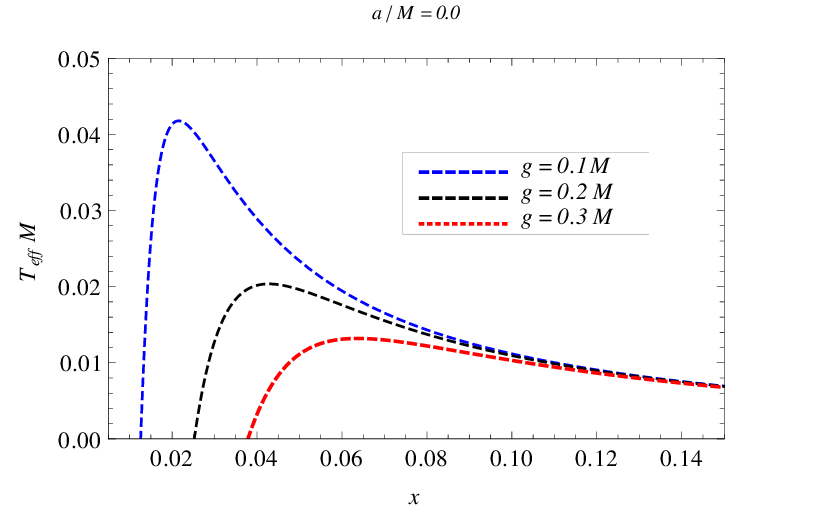}\hspace{-0.2cm}
&\includegraphics[scale=0.62]{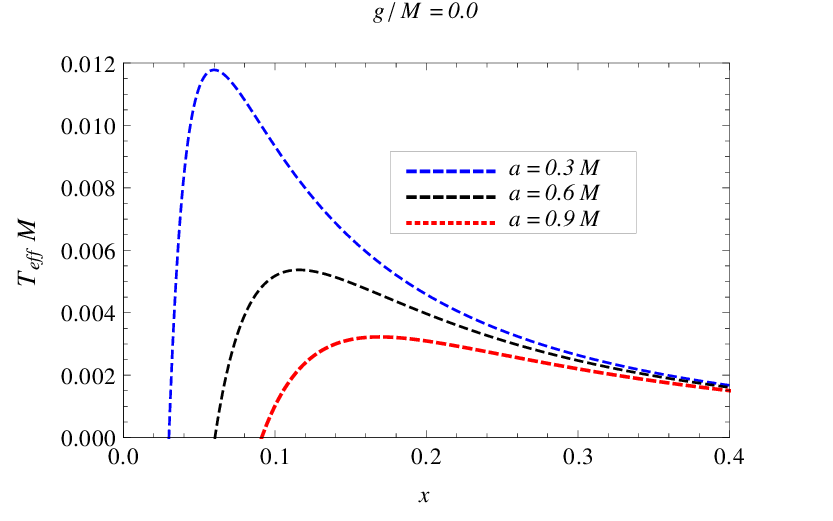}\\
\includegraphics[scale=0.62]{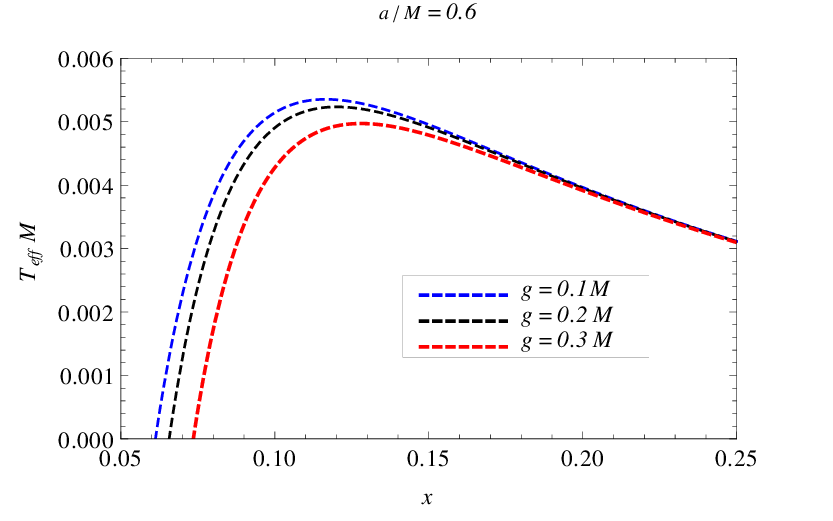}\hspace{-0.2cm}
&\includegraphics[scale=0.62]{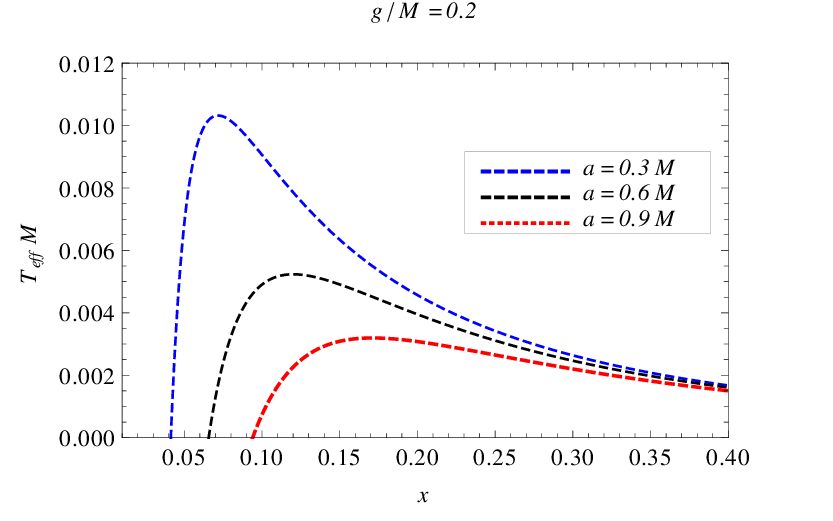}\\
\end{tabular}
\caption{The plot showing the behaviour of effective temperature $T_{eff}M$ vs $x$ for rotating regular-dS black holes and a comparison with the temperature of regular-dS black holes ($a=0$) and the Kerr-dS black holes ($g=0$).}\label{Teff2}
\end{figure}
\begin{figure}
\begin{tabular}{c c c c c}
 \includegraphics[scale=0.62]{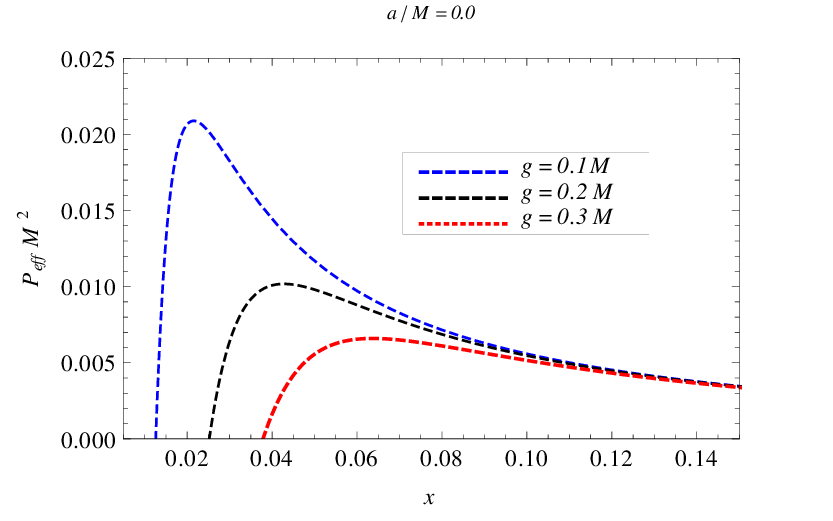}\hspace{-0.2cm}
&\includegraphics[scale=0.62]{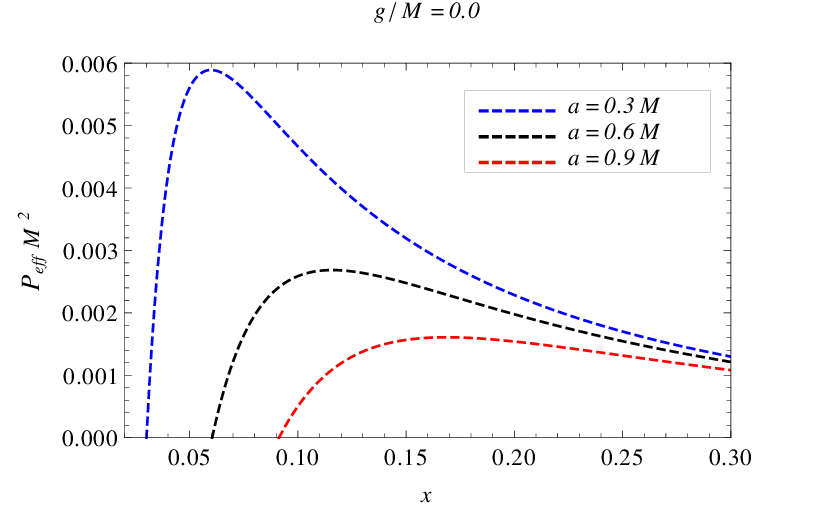}\\
\includegraphics[scale=0.62]{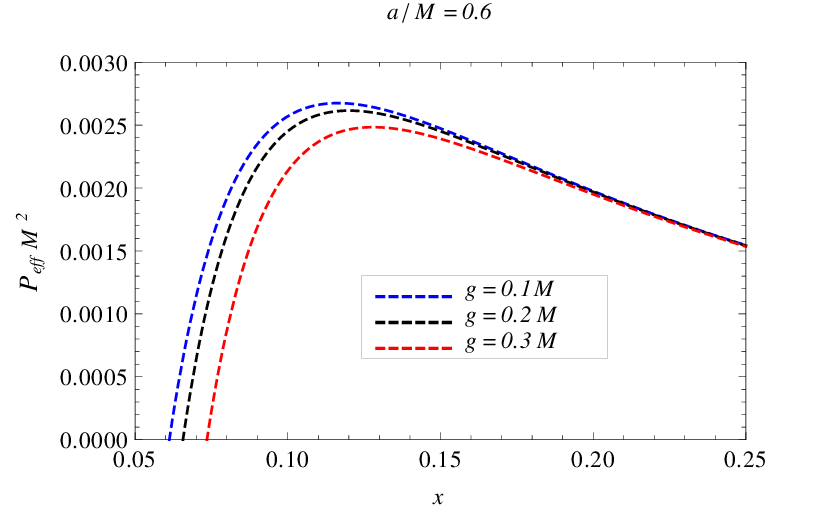}\hspace{-0.2cm}
&\includegraphics[scale=0.62]{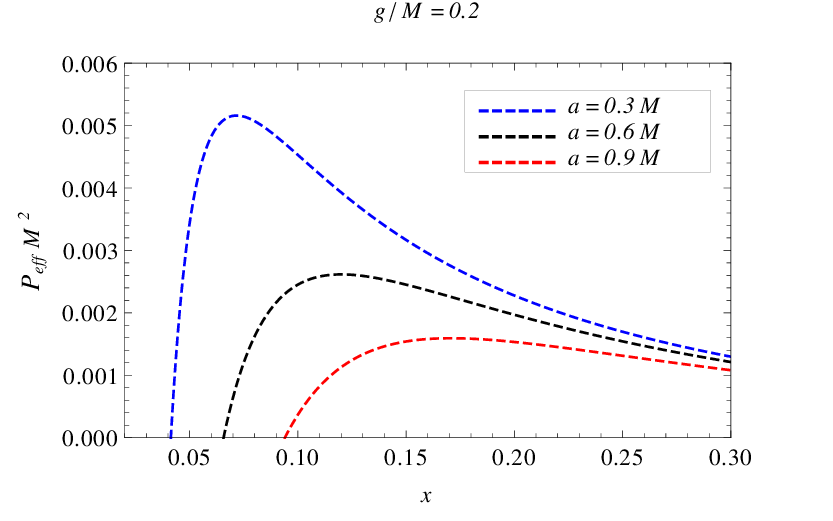}\\
\end{tabular}
\caption{The plot showing the behaviour of effective pressure $P_{eff}M^2$ vs $x$ for rotating regular-dS black holes and a comparison with the temperature of spherical regular-dS black holes ($a=0$) and the Kerr-dS black holes ($g=0$).}\label{Peff}
\end{figure}
\subsection{Phase transition}
\begin{figure}
\begin{tabular}{c c c c c}
 \includegraphics[scale=0.62]{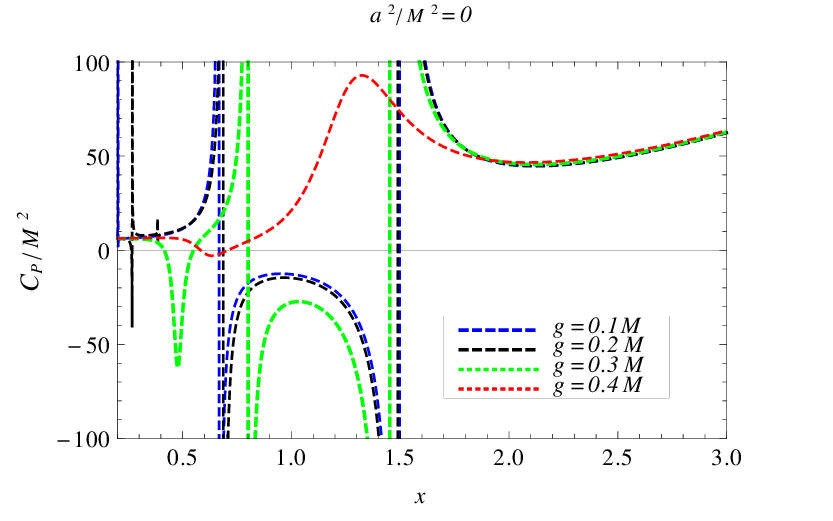}\hspace{-0.2cm}
&\includegraphics[scale=0.62]{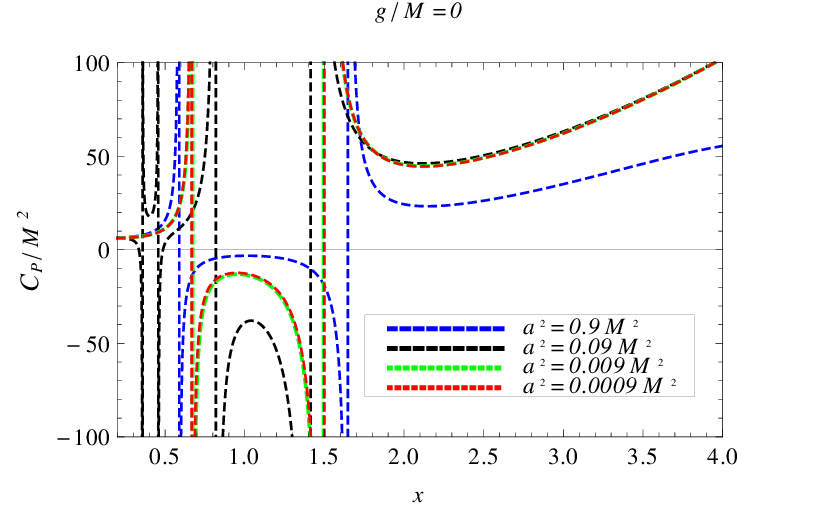}\\
\includegraphics[scale=0.62]{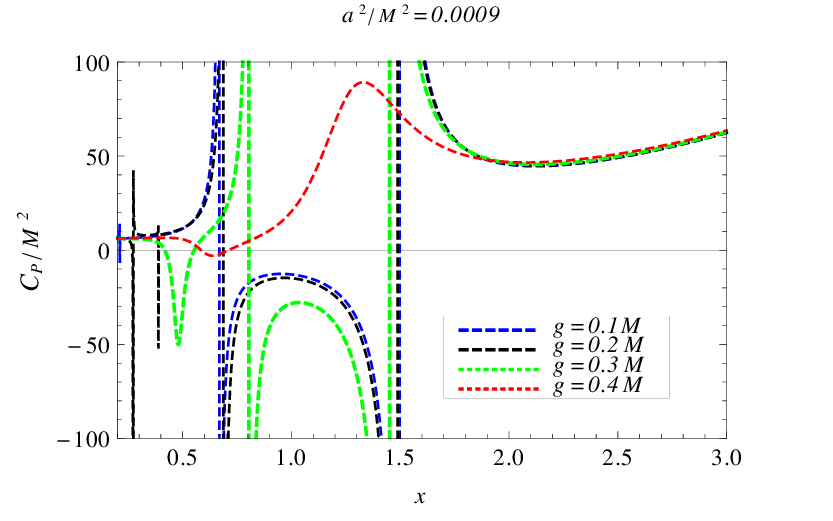}\hspace{-0.2cm}
&\includegraphics[scale=0.62]{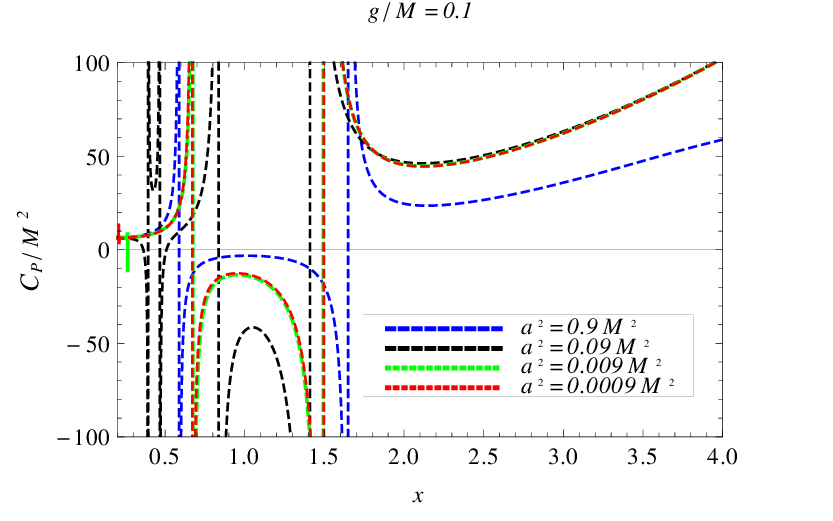}
\end{tabular}
\caption{The plot showing the behaviour of the heat capacity $C_P/M^2$ vs $x$ for rotating regular-dS black holes and a comparison with the temperature of spherical regular-dS black holes ($a=0$) and the Kerr-dS black holes ($g=0$).}\label{CP2}
\end{figure}

Next, we focus our attention on the phase transition of the   rotating regular-dS black holes by calculating the heat capacity at constant pressure $C_P$, the volume expansion coefficient $\alpha$, and the isothermal compressibility $\kappa_T$. They, respectively, are given by 
\begin{eqnarray}\label{heatcapP}
C_P&=&T_{eff}\left(\frac{\partial S}{\partial T_{eff}}\right)_{P_{eff}}=T_{eff}\Bigg[\frac{\left(\frac{\partial S}{\partial x}\right)_{r_C}\left(\frac{\partial P_{eff}}{\partial r_C}\right)_{x}-\left(\frac{\partial S}{\partial r_C}\right)_{x}\left(\frac{\partial P_{eff}}{\partial x}\right)_{r_C}}{\left(\frac{\partial T_{eff}}{\partial x}\right)_{r_C}\left(\frac{\partial P_{eff}}{\partial r_C}\right)_{x}-\left(\frac{\partial T_{eff}}{\partial r_C}\right)_{x}\left(\frac{\partial P_{eff}}{\partial x}\right)_{r_C}}\Bigg],
\end{eqnarray}
\begin{eqnarray}\label{alphaP}
\alpha &=& \frac{1}{V}\left(\frac{\partial V}{\partial T_{eff}}\right)_{P_{eff}}=\frac{1}{V}\Bigg[\frac{\left(\frac{\partial V}{\partial x}\right)_{r_C}\left(\frac{\partial P_{eff}}{\partial r_C}\right)_{x}-\left(\frac{\partial V}{\partial r_C}\right)_{x}\left(\frac{\partial P_{eff}}{\partial x}\right)_{r_C}}{\left(\frac{\partial T_{eff}}{\partial x}\right)_{r_C}\left(\frac{\partial P_{eff}}{\partial r_C}\right)_{x}-\left(\frac{\partial T_{eff}}{\partial r_C}\right)_{x}\left(\frac{\partial P_{eff}}{\partial x}\right)_{r_C}}\Bigg],
\end{eqnarray}
\begin{eqnarray}
\kappa_T &=&-\frac{1}{V}\left(\frac{\partial V}{\partial P_{eff}}\right)_{T_{eff}}=-\frac{1}{V}\Bigg[\frac{\left(\frac{\partial V}{\partial x}\right)_{r_C}\left(\frac{\partial T_{eff}}{\partial r_C}\right)_{x}-\left(\frac{\partial V}{\partial r_C}\right)_{x}\left(\frac{\partial T_{eff}}{\partial x}\right)_{r_C}}{\left(\frac{\partial T_{eff}}{\partial x}\right)_{r_C}\left(\frac{\partial P_{eff}}{\partial r_C}\right)_{x}-\left(\frac{\partial T_{eff}}{\partial r_C}\right)_{x}\left(\frac{\partial P_{eff}}{\partial x}\right)_{r_C}}\Bigg].
\end{eqnarray}
We obtain the critical values  effective pressure $P_{eff}$=$P_{eff}^c$ and effective temperature $T_{eff}$=$T_{eff}^c$ at  $x$=$x^c$, and study the critical behaviour in the extended phase space.  The critical points $x^c$ in the phase space are defined by the divergence of the heat capacity. So,  solving the denominator in Eq.~(\ref{heatcapP}), we  obtain the critical points where, incidentally, temperature $T_{eff}^c$ and pressure $P_{eff}^c$ take maximum.  The values $x^c$, $T_{eff}^c$, and $P_{eff}^c$ corresponding to $r_C=10$ are shown in the Table~\ref{crict1}. To discuss the local thermal stability of the black holes at constant pressure, we note that when the heat capacity $C_P > 0 (<0)$, the black hole is locally stable (unstable) to thermal fluctuations. The specific heat at constant pressure $P$ is given by Eq.~(\ref{heatcapP}) and it's behaviour depicted in Fig.~\ref{CP2}. The three regions in the  Fig.~\ref{CP2} correspond to small, intermediate and large black holes depending on the value of $x$, and the sign of $C_P > 0 (<0)$ in these regions determine the stability (instability), e.g., the large black holes are always stable (Fig.~\ref{CP2}).   This behaviour of $C_P$ can be related to $P$ relative the critical values $P^c_{eff}$ ( see Table.\ref{crict1}) such that for  $P>P^c_{eff}$, there is no singular points for $C_P$. 
\begin{table}[h]
    \begin{center}
        \begin{tabular}[b]{|c|c|c|c|c|c|c|c|c|c|c|c|c|c|c|c|c}
            \hline 
            &\multicolumn{3}{c}{$a=0.0$} &&\multicolumn{4}{c|}{$a=0.3$}
            &\multicolumn{4}{c|}{$a=0.95$}
            \\ \hline &$g=0.1$& 
            $g=0.2$& $g=0.3$& $g=0.4$&
            $g=0.1$&
            $g=0.2$& $g=0.3$& $g=0.4$
            &
            $g=0.1$&
            $g=0.2$& $g=0.3$& $g=0.4$
            \\ \hline & & & & & & & & & & & & 
            \\
            $x^{c}$
            &0.0215&0.0427&0.0637& 0.0843
             &0.0619&0.0717&0.08623&0.1027 
             &0.1779&0.1795&0.1834&0.1901
            \\ \hline & & & & & & & & & & & & 
            \\
            $T_{eff}^{c}$&0.0418&0.0272&0.01827&0.0135
                   &0.0115&0.0103&0.0087&0.0072 
                   &0.0029&0.0029&0.0029&0.0029
            \\ \hline & & & & & & & & & & & & 
            \\
            $P_{eff}^{c}$&0.0209&0.0136&0.0091&0.0067
 &0.0057&0.0051&0.0043&0.0036
 &0.00149&0.00148&0.00145&0.0014
            \\ \hline 
        \end{tabular}
        \caption{The tabulated values of the critical ratio $x^c$, the critical temperature $T_{eff}^c$ and the critical pressure $P_{eff}^{C}$ for different values of the charge parameter $g$ and rotation parameter $a$.\label{crict1}}
    \end{center}
\end{table}
\begin{figure}
\begin{tabular}{c c c c c}
 \includegraphics[scale=0.62]{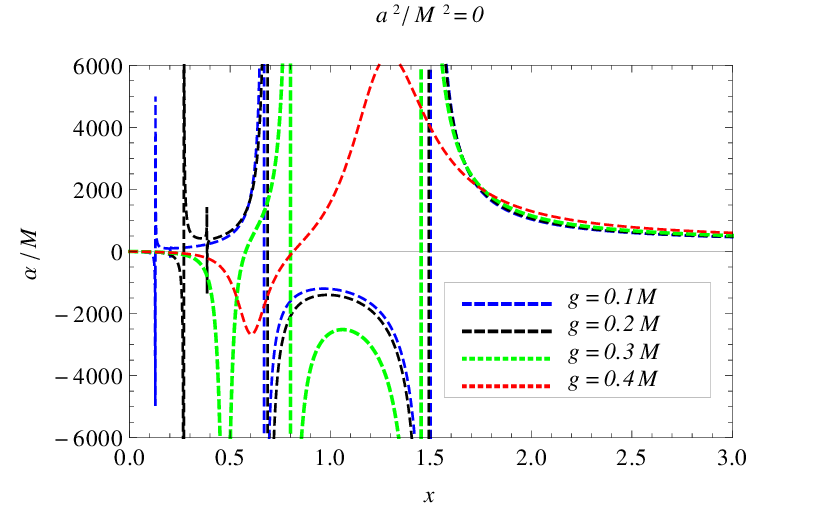}\hspace{-0.2cm}
&\includegraphics[scale=0.62]{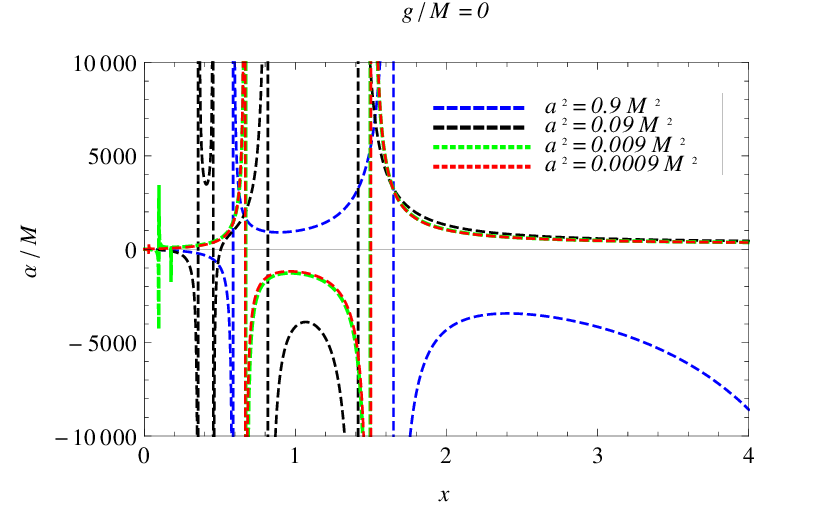}\\
\includegraphics[scale=0.62]{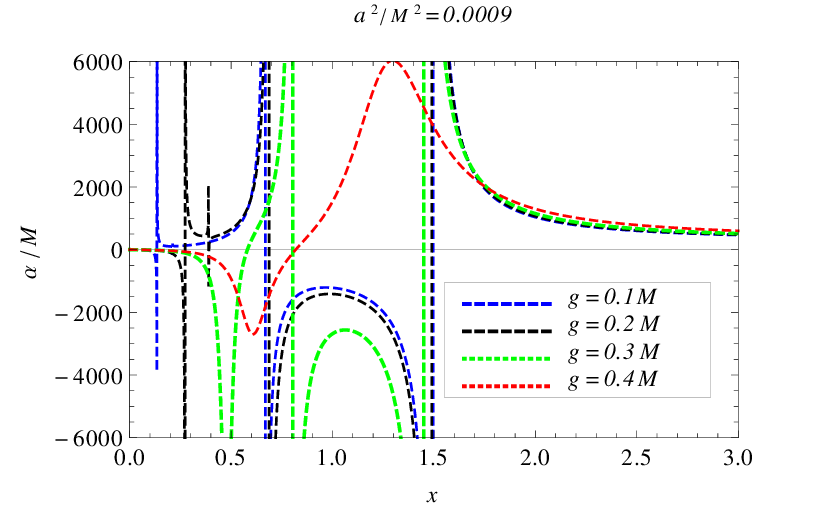}\hspace{-0.2cm}
&\includegraphics[scale=0.62]{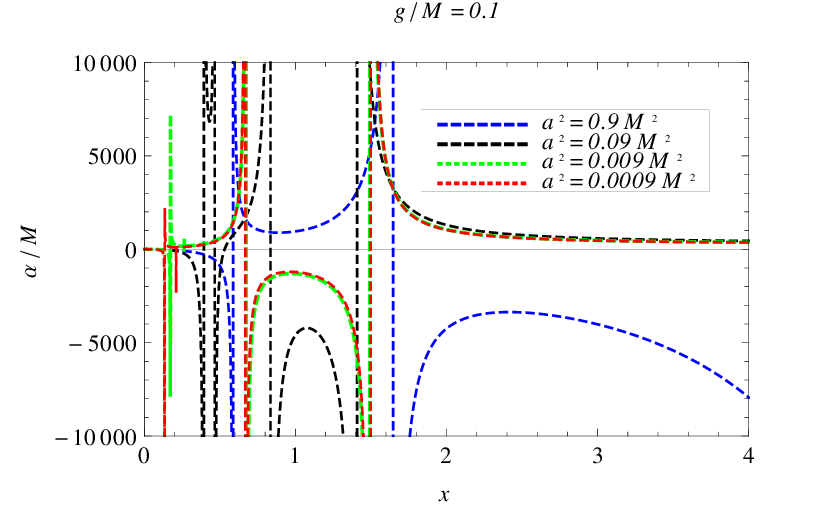}
\end{tabular}
\caption{The plot showing the behaviour of  volume expansion coeeficient $\alpha/M$ vs $x$ for rotating regular-dS black holes and a comparison with the temperature of spherical regular-dS black holes ($a=0$) and Kerr-dS black holes ($g=0$).}\label{beta2}
\end{figure}
\begin{figure}
\begin{tabular}{c c c c c}
\includegraphics[scale=0.62]{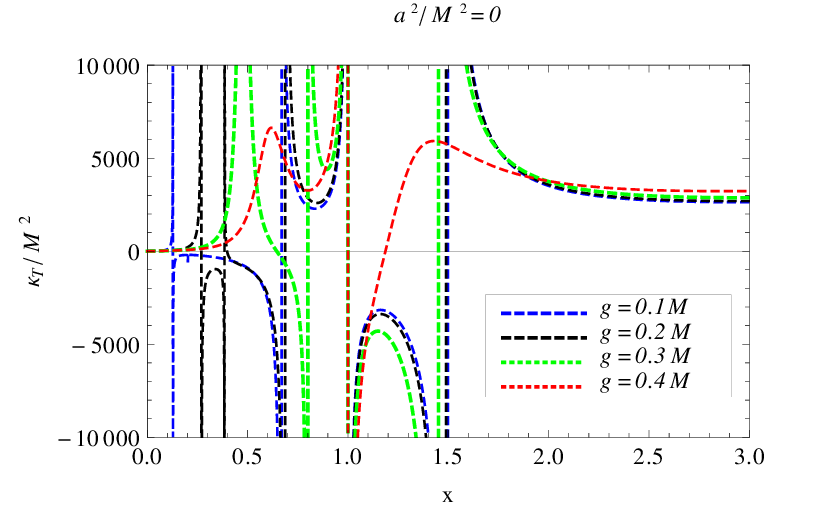}\hspace{-0.2cm}
&\includegraphics[scale=0.62]{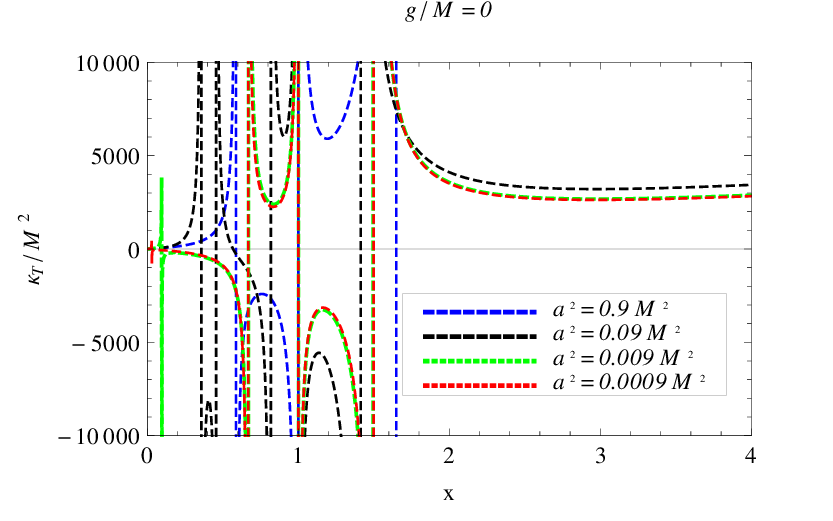}\\
 \includegraphics[scale=0.62]{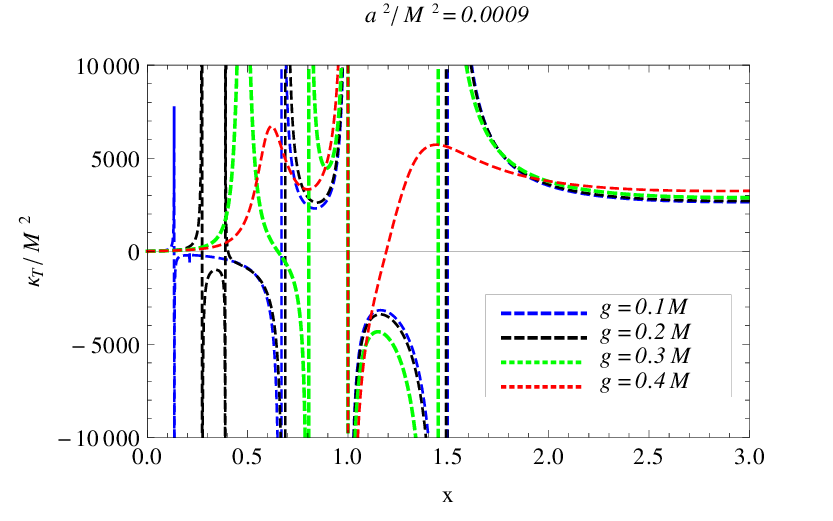}\hspace{-0.2cm}
&\includegraphics[scale=0.62]{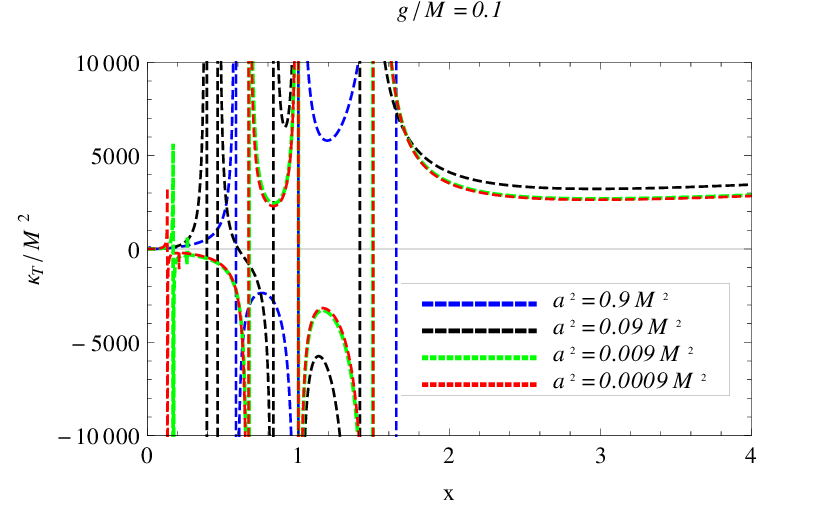}
\end{tabular}
\caption{The plot showing the behaviour of isothermal compressibility $\kappa_T/M^2$ vs $x$ for rotating regular-dS black holes with a comparison of the temperature of regular-dS black holes ($a=0$) and the Kerr-dS black holes ($g=0$).}\label{kappa2}
\end{figure}
\subsection{A corrected entropy}\label{correntropy_for}
We now proceed to calculate the corrected entropy of the rotating regular-de Sitter black holes. The total entropy of the rotating regular-de Sitter black holes are written as
\begin{eqnarray}
\label{correntropy}
S=S_++S_C+f(x)=\frac{\pi r_C^2}{\Xi}\left(1+x^2+\frac{f(x)\Xi}{r_C^2}+2\frac{a^2}{r_C^2}\right)
\approx{\pi r_C^2}\left(1+x^2+f(x)+2\frac{a^2}{r_C^2}\right)
\end{eqnarray}
In the above expression, we take $a<<l$ and $r_C<l$, so that $\Xi\to 1$. Next task is to determine the unknown function appearing in the expression (\ref{correntropy}). In these approximations, we iterate the mass term to be
\begin{eqnarray}
\label{mssterm}
&&M\approx\frac{r_C\left(1+x\right)\left(x^2+a^2/r_C^2\left(1+x^2\right)\right)\left(x^3+g^3/r_C^3\left(1+x^3\right)\right)}{2x^4\left(x^2+x+1\right)}
\end{eqnarray}
Obviously, when $g=0$, we have the mass term to be  \cite{Li:2016zdi} 
$$M\approx\frac{r_C\left(1+x\right)\left(x^2+a^2/r_C^2\left(1+x^2\right)\right)}{2x\left(x^2+x+1\right)}$$
As we take $a<<r_+$ and $\Xi\to 1$, the effective temperature of the rotating regular de Sitter black holes can be approximated as
\begin{eqnarray}
\label{teffexpress}
&&T_{eff}\approx\frac{\left(\frac{\partial{M}}{\partial{x}}\right)_{r_C,a}\left(\frac{\partial{V}}{\partial{r_C}}\right)_{x}-\left(\frac{\partial{M}}{\partial{r_c}}\right)_{x,a}\left(\frac{\partial{V}}{\partial{x}}\right)_{r_C}}{\left(\frac{\partial{S}}{\partial{x}}\right)_{r_C,a}\left(\frac{\partial{V}}{\partial{r_C}}\right)_{x}-\left(\frac{\partial{S}}{\partial{r_c}}\right)_{x,a}\left(\frac{\partial{V}}{\partial{x}}\right)_{r_C}}\nonumber\\
&=&\frac{1}{2\pi{r_C}}\frac{T_1(x)}{T_2(x)}
\end{eqnarray}
where
\begin{small}
\begin{eqnarray}
\label{T1T2}
T_1(x)&=&\frac{1}{x^2(1+x+x^2)}\Bigg[x^2(1+2x)(1-x)+x^5(1+x)-\frac{a^2}{r_C^2}\Big[(1+x+x^2)(1+x^4)-2x^3\nonumber\\
&-&\frac{2g^3}{r_C^3x^3}\left((1+x+x^2)(2+x^3+2x^7)+x^6\right)\Big]+\frac{g^3}{r_C^3x^3}\left(2x^2(1+x+x^2)(1+x^5)-2x^5(1+x)\right)\Bigg]\nonumber\\
T_2(x)&=&\left(2 x (1+x)+2 x^2 f(x)+\left(1-x^3\right) f^\prime(x)\right)
\end{eqnarray}
\end{small}
Neglecting the higher order terms in $a$, we have the two horizons temperatures as
\begin{small}
\begin{eqnarray}
\label{T1T2}
T_+(x)&=&\frac{1}{4\pi{r_C}}\frac{A_1(x)}{B_1(x)}\nonumber\\
T_C(x)&=&\frac{1}{4\pi{r_C}}\frac{A_2(x)}{B_2(x)}
\end{eqnarray}
\end{small}
where, $A_1,\;A_2,\;B_1,\;\text{and}\;B_2$ are complicated functions of $a,\;r_C,\;g,\;\text{and},\;x$ and are not written here as they are of little importance.
When the radiation temperatures of both the horizons are equal, we have
\begin{small}
\begin{eqnarray}
\label{degen1}
\frac{a^2}{r_C^2}&=&\frac{1}{D_1(x)}\Big[1+5x+5x^2+5x^3+x^4+\frac{g^3}{r_C^3x^3}(1+x) \left(2+3 x+2 x^2+2 x^3\right) \left(2+2 x+3 x^2+2 x^3\right)\nonumber\\
&-&\sqrt{1+10x+31x^2+56x^3+69x^4+56x^5+31x^6+10x^7+x^8+\frac{2g^3}{r_C^3x^3}D_2(x)}\Big]
\end{eqnarray}
\end{small}
where $
D_1(x)=2\left((x^2+1)-\frac{g^3}{r_C^3x^3}(1+x)(2-x+2x^2)\right),\;D_2(x)=(1+x) \Big(4+30 x+94 x^2+193 x^3+289 x^4+325 x^5+289 x^6$ $+193 x^7+94 x^8+30 x^9+4 x^{10}\Big)$.
These equations in the limit of $g=0$, reduces to the corresponding expressions for Kerr black holes  \cite{Li:2016zdi} $$\frac{a^2}{r_C^2}=\frac{\Big[1+5x+5x^2+5x^3+x^4-\sqrt{1+10x+31x^2+56x^3+69x^4+56x^5+31x^6+10x^7+x^8}\Big]}{2(x^2+1)}$$
In general, the horizon temperatures of the black hole event horizon and the cosmological horizon are not equal, hence we cannot compare them with the effective temperature. But, in particular case, such as the lukewarm black holes do have the same temperature for the two horizons. In such case, we conjecture that the effective temperature also has the equal value to that of the horizon temperatures. Once this condition is met, we can have the information about the unknown function $f(x)$. For the rotating regular-de Sitter black holes
\begin{eqnarray}
\label{tildeeff}
T_{eff}=\tilde{T_+}=\tilde{T_C}
\end{eqnarray}
Combining Eqs.~(\ref{teffexpress}) and (\ref{tildeeff}), we have
\begin{eqnarray}
\label{tildeeff1}
T_{eff}=\frac{T_1(x)}{\tilde{T_1}(x)}\tilde{T_+}(\tilde{T_C})
\end{eqnarray}
where $\tilde{T}_1(x)$ is the value of $T_1(x)$ with $a^2/r_C^2$ given in (\ref{degen1}). Now, substituting Eqs.~(\ref{teffexpress}) and (\ref{tildeeff}) in Eq.~(\ref{tildeeff1}), we have 
\begin{eqnarray}
\label{tildeeff1}
T_2(x)=\frac{1}{2\pi r_C}\frac{\tilde{T_1}(x)}{\tilde{T_+}(\tilde{T_C})}.
\end{eqnarray}
Substituting the expressions for $\tilde{T_1}(x)$ and $\tilde{T_C})$ from Eqs.~(\ref{teffexpress}) and (\ref{T1T2}), respectively, we have from Eq.~(\ref{tildeeff1}), a differential equation involving the function $f(x)$ and its derivative.
\begin{figure}
\label{correctionfunc}
\includegraphics[scale=0.65]{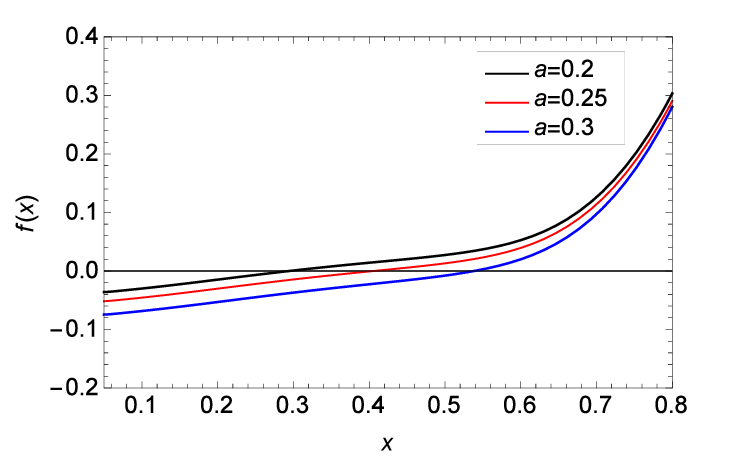}\hspace{-0.2cm} 
\includegraphics[scale=0.8]{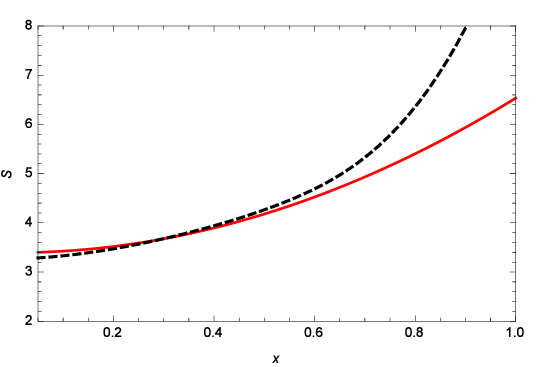} 
\caption{(Left) The correction function $f(x)$ with respect to $x$. (Right) The entropy $S$ with respect to $x$. The nonlinear charge parameter is taken to be $g=0.02$ and we set $r_C=1$. The solid red line in the right figure represents the entropy of sum of the two horizons and dashed line corresponds to the entropy given in Eq. (\ref{correntropy}).}
\end{figure}
This functional expression is not solvable exactly. In order to see the nature of the correction term $f(x)$, we have to plot them. We find that the function $f(x)$ behaves monotonically which leads to the divergence when $x\to 1$. Accordingly, we can see the behaviour of the corrected entropy once we solve the function $f(x)$ numerically. 
It is very clear to note we recover all the expressions of the thermodynamic quantities, e.g., $M$, $T_{eff}$, $\tilde{T_C(\tilde{T_+})}$ of the Kerr black holes when $g=0$.  \cite{Li:2016zdi}
\section{Conclusions}\label{conclusion}
The recent year witnessed an increasing interest in the study of rotating regular black holes, which depend on an additional deviation parameter coming from NED that measures potential deviations from the Kerr black holes. Further, although the astrophysical black holes are supposed to be the Kerr black hole, the actual nature of the astrophysical black hole still needs to be tested. The rotating regular black hole is one such black hole that may be suitable for an astrophysical black hole. Because of this, we have performed a comprehensive analysis of thermodynamics for rotating regular black holes to calculate the thermodynamic quantities such as the Hawking temperature, heat capacity, and thermal behaviour. Interestingly, the rotating regular black holes admit a thermodynamically stable double-horizon remnant, $M=M_{+}^C$, corresponding to the minimal mass configuration. At the point $r_+=r_+^C$, the heat capacity diverges and thereby testifying a quantum cooling and a second-order phase transition during evaporation \cite{Dymnikova:1996zz, Dymnikova:2010zz}.

There are intensive studies on the black hole thermodynamics in either asymptotically flat and asymptotically AdS holes, the latter emerging because of the AdS/CFT correspondence. However, the thermodynamics of asymptotically dS black holes possesses difficulties and remains relatively unexplored as de Sitter black hole spacetimes are essentially non-equilibrium because the event and cosmological horizons are marked by its temperature. Next, we extended the rotating regular black holes in dS spacetimes and characterized the resulting physical quantities by its horizon structure which can be at the most three. We find two critical masses $M_{\text{cr1}}\;\text{and}\;M_{\text{cr2}}$ such that  when $M_{\text{cr1}}<M<M_{\text{cr2}}$, we have three distinct horizons. It is demonstrated that  $M=M_{\text{cr1}}$ corresponds to the degenerate Cauchy and event horizons $r_-=r_+=r_+^E$ whereas when $M=M_{\text{cr2}}$, we have the degenerate event and cosmological horizons $r_+=r_C=r_C^E$.  We have discussed the thermodynamics of rotating regular black holes by treating the cosmological constant as thermodynamic pressure and its conjugate quantity as thermodynamic volume. We also formulate separate thermodynamic first laws for each horizon present in the dS spacetime and study their thermodynamics as if they were independent systems characterized by their temperature. When the two horizons, the event horizon and the cosmological horizon, coincide, one gets the Nariai or lukewarm black holes which is in thermodynamic equilibrium. We have presented the globally thermodynamic properties, including thermal stability and phase transition of rotating regular-dS black holes. The connection, namely $x$, between the event horizon and the cosmological horizon is introduced to calculate the effective thermodynamic quantities of the rotating regular-dS black holes. Finally, we have calculated the globally effective thermodynamic quantities such as the effective temperature, the effective pressure, the heat capacity at constant pressure, the isothermal compressibility and the volume expansion coefficient to deal with the rotating regular-dS black holes. Indeed, the rotating regular-dS black holes are relatively cooler when compared with the Kerr-dS black hole. Finally, we have dealt with the corrected entropy obtained when adding the correction function and the entropies of the event and cosmological horizons. We find that such correction would significantly affect a large distance limit. 

Finally, considering the relation between the event horizon and the cosmological horizon and their thermodynamic interplay, we conjecture that the total entropy of the regular de Sitter black hole must have a corrected function in an additive way. We found that the entropy of the rotating regular de Sitter black holes is ever increasing with the increase in the value of $x$. As $x\to 1$, we found that it is exceedingly significant.
 
\acknowledgements
S.G.G. and M.S.A. thank DST INDO-South Africa (INDO-SA) bilateral project DST/INT/South Africa/P-06/2016 and also thanks University of KwaZulu-Natal, Astrophysics and Cosmology Research Unit (ACRU), Durban, South Africa, for the hospitality while this work was being done. The research of M.~S.~A. is supported by the National Postdoctoral Fellowship of the Science and Engineering Research Board (SERB), Department of Science and Technology (DST), Government of India, File No., PDF/2021/003491.

\end{document}